%% file: ms.tex
\documentclass[floatfix,amsmath,amssymb,prl,aps,twocolumn,mathbbm, reprint, superscriptaddress]{revtex4-2}

\usepackage{pdfpages}
\usepackage{pgffor}
\usepackage{amsmath,amssymb}
\usepackage{graphicx}
\usepackage{csquotes}
\usepackage{xcolor}
\usepackage{mathtools}
\usepackage{physics}
\usepackage[alsoload=synchem,alsoload=abbr,binary-units]{siunitx}
\usepackage[inline]{enumitem}
\usepackage{hyperref}

\makeatletter
\AtBeginDocument{\let\LS@rot\@undefined}
\makeatother

\hypersetup{
    colorlinks,
    linkcolor={red!50!black},
    citecolor={blue!50!black},
    urlcolor={blue!80!black}
}

\newcommand{\fleft}{\mathopen{}\mathclose\bgroup\left}
\newcommand{\fright}{\aftergroup\egroup\right}

\graphicspath{{./figs/}}

\newcommand{\sectionstart}{\hspace{-1.9ex}---}

\begin{document}

\title{Long-Range Non-Equilibrium Coherent Tunneling Induced by Fractional Vibronic Resonances} 

\author{R.~Kevin~Kessing}
\email{kevin.kessing@uni-ulm.de}
\affiliation{Institut f\"ur Theoretische Physik, Universit\"at Ulm, 89069 Ulm, Germany} 
\affiliation{Institut f\"{u}r Theoretische Physik, Georg-August-Universit\"{a}t G\"{o}ttingen, 37077 G\"{o}ttingen, Germany}
\affiliation{Department of Chemistry, Massachusetts Institute of Technology, Cambridge, Massachusetts 02139}
\author{Pei-Yun Yang}
\email{pyyang@phys.ntu.edu.tw}
\affiliation{Department of Physics, National Taiwan University, Taipei 10617, Taiwan (R.O.C.)}
\affiliation{Beijing Computational Science Research Center, Beijing 100193, China}
\affiliation{Department of Chemistry, Massachusetts Institute of Technology, Cambridge, Massachusetts 02139}
\author{Salvatore R.~Manmana}
\email{salvatore.manmana@uni-goettingen.de}
\affiliation{Institut f\"{u}r Theoretische Physik, Georg-August-Universit\"{a}t G\"{o}ttingen, 37077 G\"{o}ttingen, Germany}
\affiliation{Fachbereich Physik, Philipps-Universit\"{a}t Marburg, 35032 Marburg, Germany} 
\author{Jianshu Cao}
\email{jianshu@mit.edu}
\affiliation{Department of Chemistry, Massachusetts Institute of Technology, Cambridge, Massachusetts 02139}

\begin{abstract}
We study the influence of a linear energy bias on a non-equilibrium excitation on a chain of molecules coupled to local phonons (a tilted Holstein model)
using both a random-walk rate kernel theory and a nonperturbative, massively parallelized adaptive-basis algorithm. 
We uncover structured and discrete vibronic resonance behavior fundamentally different from both linear response theory and homogeneous polaron dynamics. 
Remarkably, resonance between the phonon energy $\hbar\omega$ and the bias $\delta_\epsilon$ occurs not only at integer but also fractional ratios $\delta_\epsilon/(\hbar\omega) = \frac{m}{n}$, which effect long-range $n$-bond $m$-phonon tunneling. 
These observations are also reproduced in a model calculation of a recently demonstrated Cy3 system.
Potential applications range from molecular electronics to optical lattices and artificial light harvesting via vibronic engineering of coherent quantum transport.
\end{abstract}

\maketitle

\input{main_contents}

\section{ACKNOWLEDGMENTS}
We thank Fabian Heidrich-Meisner, Martin Plenio and Andrea Mattioni for insightful discussions and feedback, as well as Benedikt Kloss and Gabriela Schlau-Cohen for providing access to their data.

RKK acknowledges generous scholarships and travel funds provided by the Studienstiftung des deutschen Volkes.
PYY acknowledges support from National Natural Science Foundation of China (Grant No. U1930402).
RKK and SRM acknowledge 
	funding by the Deutsche Forschungsgemeinschaft (DFG, German Research Foundation) -- 217133147/SFB 1073, project B03.
JC acknowledges support from the NSF (Grants No. CHE 1800301 and No. CHE 1836913) and MIT Sloan Fund.

RKK acknowledges computational resources provided by the state of Baden-W\"urttemberg through bwHPC
and the DFG through grant No.~INST~40/575-1~FUGG (JUSTUS~2 cluster), 
and gratefully acknowledges GPU resources provided by the Institute of Theoretical Physics in G\"ottingen and financed by the DFG and the Bundesministerium f\"ur Bildung und Foschung (BMBF).

\input{ms.bbl}


\section*{Supplementary material}

# Supporting Information for:
# Long-Range Non-Equilibrium Coherent Tunneling Induced by Fractional Vibronic Resonances

R. Kevin Kessing,[1, 2, 3, *] Pei-Yun Yang,[4, 5, 3, †] Salvatore R. Manmana,[2, 6, ‡] and Jianshu Cao[3, §]

[1] Institut für Theoretische Physik, Universität Ulm, 89069 Ulm, Germany
[2] Institut für Theoretische Physik, Georg-August-Universität Göttingen, 37077 Göttingen, Germany
[3] Department of Chemistry, Massachusetts Institute of Technology, Cambridge, Massachusetts 02139
[4] Department of Physics, National Taiwan University, Taipei 10617, Taiwan (R.O.C.)
[5] Beijing Computational Science Research Center, Beijing 100193, China
[6] Fachbereich Physik, Philipps-Universität Marburg, 35032 Marburg, Germany

(Dated: May 12, 2022)

## CONTENTS

## 1. NUMERICAL METHOD: PARMUDA

### 1.a. Motivation: Most basis states contribute negligibly

The ParMuDA method is an extension of a simple matrix-vector time evolution algorithm in which the "relevant" parts of the Hilbert space are adaptively added and removed. The motivation for this is that most of a Hilbert space is irrelevant to a calculation[1]; however, the parts of the Hilbert space that are most relevant at one point in time may not be the most relevant at another point in time.

A simple illustrative example is a Gaussian state in a harmonic trap (directly equivalent to the phonons under consideration): In the energy eigenbasis, the Gaussian state at $(q = 0, p = 0)$ is exactly described by a single eigenstate. However, upon displacing the state to a different $(q, p)$, this initial eigenstate may no longer be appreciably occupied, and instead the occupation will shift mostly to another set of eigenstates. This principle of

* kevin.kessing@uni-ulm.de
† pyyang@phys.ntu.edu.tw
‡ salvatore.manmana@uni-goettingen.de
§ jianshu@mit.edu

following the basis states occupied by the state vector at each given time is the basic idea underlying ParMuDA. We find in the actual Holstein calculations that less than half of the expansion coefficients account for 99.99 % of the total weight of the state vector. Assuming the distribution of the expansion coefficients to be exponential, we could therefore keep only $n$ basis states and neglect the remaining basis states and incur an error only of order $\mathcal{O}(e^{-\lambda n})$.

### 1.b. Overview of the method

The basic idea of the presently used method is inspired by the idea underlying tDMRG: To efficiently truncate the high-dimensional Hilbert space into a smaller Hilbert space in a time-dependent manner[2;3]. However, while DMRG methods are based on singular value decompositions of the local matrices to determine the effective Hilbert space, the present method uses a distinct weighting and adaptation algorithm to construct effective Hilbert space and state vectors.

Given a Hilbert space with an arbitrary countable orthonormal basis set $\{|n\rangle\}$, the basic algorithm of a single time evolution step is:

**Input:** Start with a given state vector (= wavefunction) $|\psi(t_0)\rangle$ which has non-zero basis coefficients $\psi_n = \langle n|\psi\rangle$ for a limited number of basis states $|n\rangle$, then:

1. *construct a Hilbert subspace*, or "effective Hilbert space", $\mathcal{H}_{eff}$ of the entire Hilbert space; this Hilbert subspace should capture both

   (a) the current state vector,

   (b) and "neighboring" basis states into which the state vector may evolve;

2. construct the Hamiltonian matrix acting on $\mathcal{H}_{eff}$ and *evolve* the state vector $|\psi(t)\rangle$ using the subspace Hamiltonian;

3. if necessary, *truncate* the resulting state vector by weight to retain only the $N$ most important basis states.

**Output:** the new time-evolved wavefunction $|\psi(t_0 + \delta t)\rangle$. Repeat from step 1 until the desired time has been reached.

According to its defining features, we call this method the **par**allelized **mu**lti-step-**d**ynamically **a**dapted basis set method, or ParMuDA. Below, we elaborate on some key details of these features. As the name suggests, each of these steps is tailored towards easy numerical parallelization, as most of the computational steps consist of linear algebra in the form of sparse matrix-vector multiplication, which results in a highly time-efficient algorithm that was implemented on graphics processing units (GPU)[4].

*The effective Hilbert space.* The dynamically adapted effective Hilbert space $\mathcal{H}_{\text{eff}}(t)$ is constructed as the span of a select set of basis states. These basis states consist of all of the basis states that comprise the current state vector as well as "neighboring" basis states. The neighboring basis states of the state $|\psi(t)\rangle$ are defined via the image of the Hamiltonian matrix acting on $|\psi(t)\rangle$: Given a Hamiltonian $H$, a total Hilbert space $\mathcal{H} \supseteq \mathcal{H}_{\text{eff}}$ spanned by a basis $\{|n\rangle\}$ and a state $|\psi(t)\rangle$, we define a minimal set of basis states $\mathcal{M}(t)$ supporting the state $|\psi(t)\rangle$, i.e.,

$$\mathcal{M}(t) := \text{supp}\{|\psi(t)\rangle\} = \{|n\rangle \in \mathcal{H} \mid \langle n|\psi(t)\rangle \neq 0\}.$$

Then the set of $k$-th degree neighboring states of $\mathcal{M}$ is defined as

$$\mathcal{N}_{\mathcal{M}}^{(k)} := \{|n\rangle \in \mathcal{H} \mid \langle n|H^k|m\rangle \neq 0 \,\forall\, |m\rangle \in \mathcal{M}\}.$$

We use $k = 2$ for the dynamically adapted Hilbert space in our calculations, i.e., we use

$$\mathcal{H}_{\text{eff}}(t + \delta t) = \text{span}\Big\{\mathcal{M}(t) \cup \mathcal{N}_{\mathcal{M}(t)}^{(1)} \cup \mathcal{N}_{\mathcal{M}(t)}^{(2)}\Big\}.$$

*Restriction of the Hamiltonian.* The Hamiltonian matrix is explicitly constructed via restriction to $\mathcal{H}_{\text{eff}}$. That is, the effective Hamiltonian is given by matrix elements

$$\langle m|H|n\rangle \,\forall\, m, n \in \mathcal{H}_{\text{eff}}(t).$$

This means that all matrix elements that map to or from a basis state not contained within $\mathcal{H}_{\text{eff}}(t)$ are discarded. Due to the nature of the Hamiltonian in question (which is typical of condensed-matter Hamiltonians), we are able to employ a highly efficient sparse matrix representation of these effective Hamiltonians. This sparse representation is also exploited in the time evolution, which is computed as the action of a Taylor series expansion (expanded until converged to machine precision) acting on the truncated state vector.

## 1.c. Relation to existing methods

The present method belongs broadly to the family of adaptive basis set methods, which deal with the exponential growth of the Hilbert space by choosing a small, truncated Hilbert space and systematically adapting the basis set to the evolution of the state vector.

A variety of adaptive basis set methods have been developed over the last decades, mostly in quantum chemistry. Many of these employ moving Gaussian basis sets, e.g., combined with Ehrenfest dynamics or a variational ansatz (cf. [5;6;7;8;9;10;11]). However, unlike these approaches, our ansatz employs *stationary* basis states that are adaptively added or removed. Two other methods that use adaptive stationary basis sets are the adapted trajectory-guided (aTG) scheme[12], first introduced by Bončca et al.[16]. The ansatz in LFS is also to determine an effective Hilbert space via the $n$-fold application of the Hamiltonian to a given quantum state. However, our present method is, to our knowledge, the first method to apply this principle dynamically, thereby continuously recalculating the effective Hilbert space.

## 2. RESONANCE IS BETWEEN $\delta_\epsilon$ AND $\hbar\omega$

Since we have four different Hamiltonian parameters $(\omega, J, g$ and $\delta_\epsilon)$, we demonstrate here that the resonance is only between $\delta_\epsilon$ and $\hbar\omega$, whereas changing the value of $t$ and $g$ does not shift the position of the resonance peaks. Recall that the main numerical results in the main text were obtained for $g = 4J$ and $J = \hbar\omega$ ($\delta_\epsilon$ was varied to probe the resonance behavior). For purposes of brevity, set $\hbar\omega$ in this configuration to the unit energy, i.e., the reference parameters are $J = \hbar\omega = 1$ and $g = 4$. We then vary each of $g$, $J$ and $\omega$ independently and keep the remaining parameters fixed to the aforementioned values: In Fig. S2.1, we set $J = 0.85$; in Fig. S2.2, we set $g = 3$; in Fig. S2.3, we set $\hbar\omega = 1.25$. These should be compared to Fig. 2 (right side) in the main text. All of the results shown here use only the numerical ParMuDA method and are not based on perturbation theory.

---

[1] with the exception of the later variant of PRODG[14], which used "interpolating Gaussians" that are also orthogonal

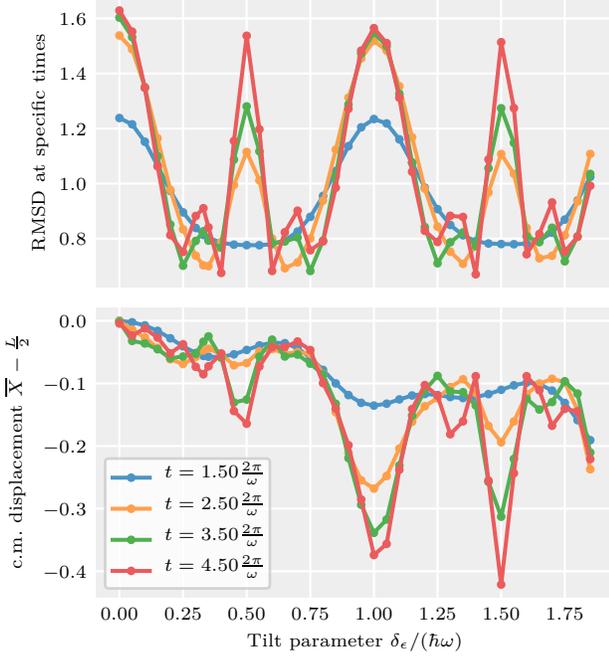

FIG. S2.1 The RMSD and center-of-mass deviation, but with $J = 0.85$ instead of $J = 1$. We see that the position of the peaks is unaffected by this change. The remaining parameters are the same: $L = 9$, $\hbar\omega = 1$, $g = 4$.

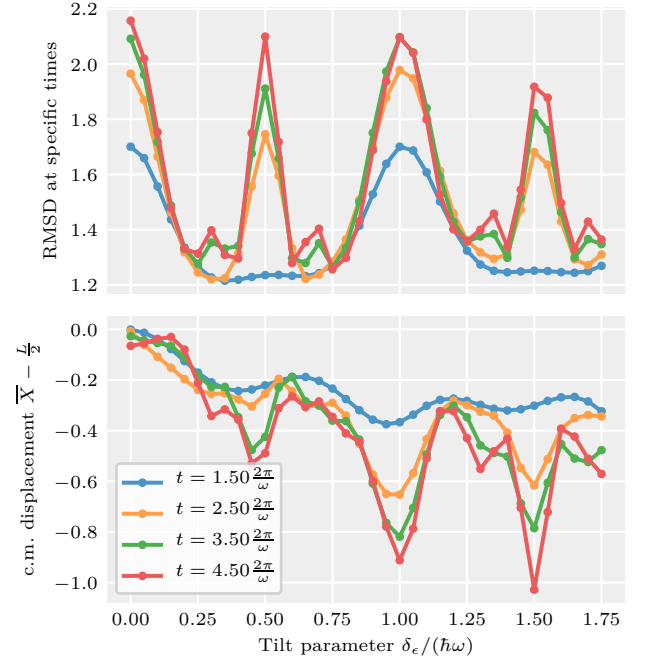

FIG. S2.2 The RMSD and average position with $g = 3$ instead of $g = 4$. Due to the decreased polaronic trapping, $L = 15$ was chosen here. The remaining parameters are the same: $J = \hbar\omega = 1$.

We see in all three plots that the position of the resonance peaks as a function of $\delta_\epsilon/(\hbar\omega)$ is unchanged. The intensity of the peaks, however, is changed, which is expected due to the effect of changing the remaining parameters: Decreasing $J$ decreases the overall particle mobility and especially the probability of long-range tunneling events; decreasing the exciton–phonon coupling $g$ decreases polaronic trapping effects and increases overall mobility; and increasing $\omega$ decreases the displacement between the excited and ground state PES minima, also decreasing polaronic effects and slightly increasing mobility (equivalently, increasing $\omega$ increases the energy of many–phonon states, making them less unfavorable and thus decreasing polaronic behavior).

## 3. GAPS OF LENGTH $n$ OCCUR FOR $\delta_\epsilon = \frac{m}{n}\hbar\omega$

In the main text, we described the appearance of local population spikes at sites $C \pm n$ for $\delta_\epsilon = \frac{m}{n}\hbar\omega$, where site $C$ is the initial and central site—in other words, we proposed that the higher-order (fractional) resonance peaks are due to tunneling events over more than one site, where the denominator of the energy ratio $\frac{\delta_\epsilon}{\hbar\omega}$ dictates the tunneling length. We shall present additional information supporting this claim by systematically investigating the local populations at sites $C - 2$ and $C - 3$, where our hypothesis proposes we should find density spikes for $\frac{\delta_\epsilon}{\hbar\omega} = \frac{m}{2}$ and $\frac{\delta_\epsilon}{\hbar\omega} = \frac{m}{3}$, respectively, with $m \in \mathbb{N}$. This

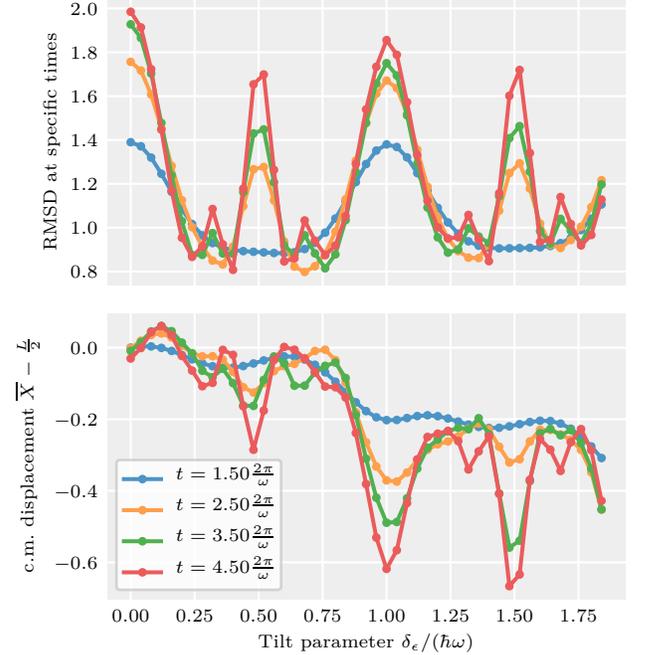

FIG. S2.3 The RMSD and average position with $\hbar\omega = 1.25$ instead of $\hbar\omega = 1$. The position of the peaks as shown here is unchanged because the $x$ axis shows $\delta_\epsilon/(\hbar\omega)$; in absolute terms, the position of the resonant values of $\delta_\epsilon$ is shifted according to the change in $\hbar\omega$. As in Fig. S2.2, $L = 15$ to accommodate smaller polaronic effects. $J = 1$, $g = 4$.

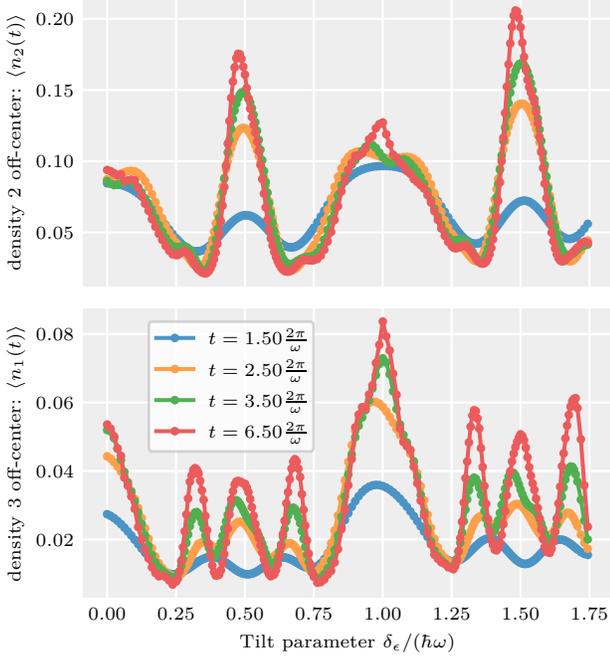

FIG. S3.1 The same calculations as in Fig. 2 (right side) of the main text, but instead of the excitonic RMSD, the local exciton density 2 sites off-center (site 2, top) and 3 sites off-center (site 1, bottom) is shown. $L = 9$, $J = \hbar\omega = 1$, $g = 4$. Data shown is numerics with interpolation (no perturbative results).

analysis reveals more information than the RMSD alone: For example, an increase in RMSD cannot differentiate between strong build-up of density at sites $C \pm 2$ or weaker build-up at sites $C \pm 3$. The change in local densities is shown in Fig. S3.1.

We see that the data supports our hypothesis and that the local density at sites $C - m$ is maximized when $\delta_\epsilon$ is an integer multiple of the $\frac{\hbar\omega}{m}$. In particular, we see that the third-order resonance peaks at $\frac{\delta_\epsilon}{\hbar\omega} = \frac{m}{3}$ present as dips in the local density at $C - 2$, and that their increased RMSD is due to build-up at $C - 3$. On the other hand, the apparent increase in density at $C - 3$ for $\frac{\delta_\epsilon}{\hbar\omega} = \frac{m}{2}$ is attributable to the tail of the strong local density spikes at $C - 2$ (notice the difference in absolute magnitudes, cf. Fig. 3 of the main text). The same also applies to $\frac{\delta_\epsilon}{\hbar\omega} \in \mathbb{Z}$, which enhances tunneling across every bond.

## 4. NUMERICS SHOW SECOND-ORDER TUNNELING FOR SMALL $J$

In the main text, we introduced a first-order (in the tunneling $J$) perturbation theory (PT) approach that provides excellent agreement with the numerical reults which take into account all orders of of the hopping operator. This agreement is valid for small values of $J$, which mostly precludes the occurrence of higher-order

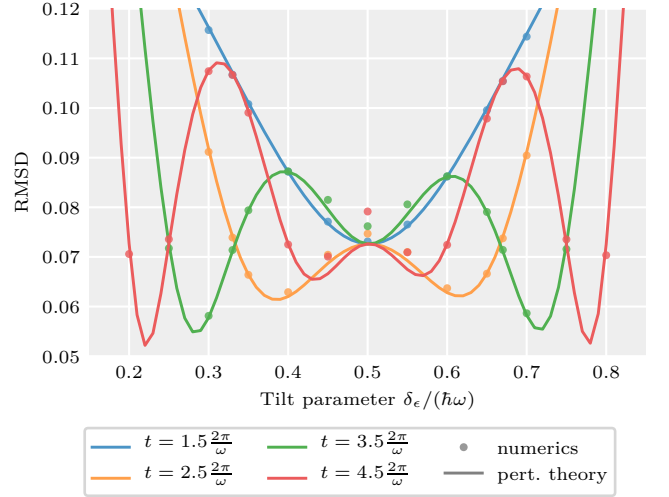

FIG. S4.1 Zoom-in on the top left of Fig. 2 from the main text: A comparison between first-order perturbation theory and numerical results for $L = 9$, $J = \frac{1}{10}\hbar\omega$, $g = 4\hbar\omega$, showing the build-up of a second-order tunneling event only in the numerics.

tunneling events which are not captured by a first-order perturbative approach.

Our hypothesis for the resonant tunneling mechanism postulates that the resonances at fractional values of the energy ratio $\frac{\delta_\epsilon}{\hbar\omega} = \frac{m}{n}$, the resonant tunneling transitions are due to $m$-phonon and $n$-th order tunneling processes. Therefore, to further validate our hypothesis, we investigate whether we observe an increase in RMSD for $\frac{\delta_\epsilon}{\hbar\omega} = \frac{1}{2}$, corresponding to second-order tunneling events, in Fig. S4.1. This Figure shows, as expected, that the overall excellent agreement for the results is slightly less good around specifically the point $\frac{\delta_\epsilon}{\hbar\omega} = \frac{1}{2}$, where the numerics provide slightly higher RMSD values than the first-order PT. This is in line with the hypothesis that $\frac{\delta_\epsilon}{\hbar\omega} = \frac{1}{2}$ enables resonant second-order tunneling processes, which are consequently captured by the numerics, but not by a first-order PT.

## 5. LINEAR RESPONSE FOR SMALL $\delta_\epsilon$

In the bottom left of Fig. 2 of the main text, we displayed the mobility, i.e., the displacement of the center of mass $\overline{X}$ from the initial position. For small values of $\delta_\epsilon$, the response of the system to a weak perturbative external gradient can be modeled as linear response, $\overline{X} = c\delta_\epsilon$. We show in Fig. S5.1 that our numerical and analytical findings also exhibit this linear behavior around the origin, with better agreement with linear response for smaller times $t$ and a smaller range of agreement for longer times. Furthermore, the slopes of the linear responses ($c$ in the above equation) are proportional to $t^2$, in agreement with Eq. 4 of the main text.

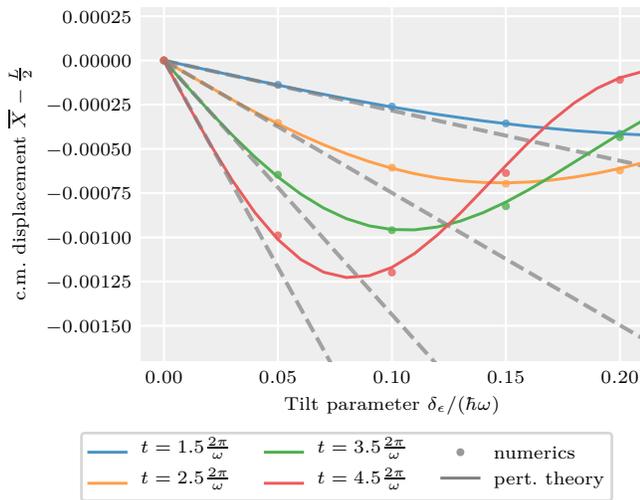

FIG. S5.1 Zoom-in on the bottom left of Fig. 2 from the main text. The dashed gray lines are extrapolations of the slope at the origin and correspond to a linear response, which is approximately valid for small times $t$ and small gradient shifts $\delta_\epsilon$. The parameters are the same as in Fig. S4.1: $L = 9$, $J = \frac{1}{10}\hbar\omega$, $g = 4\hbar\omega$.

## REFERENCES

(1) Verstraete, F.; Murg, V.; Cirac, J. Matrix Product States, Projected Entangled Pair States, and Variational Renormalization Group Methods for Quantum Spin Systems. *Adv. Phys.* **2008**, *57*, 143–224.

(2) Manmana, S. R.; Muramatsu, A.; Noack, R. M. Time Evolution of One-Dimensional Quantum Many Body Systems. *AIP Conf. Proc.* **2005**, *789*, 269–278.

(3) Paeckel, S.; Köhler, T.; Swoboda, A.; Manmana, S. R.; Schollwöck, U.; Hubig, C. Time-Evolution Methods for Matrix-Product States. *Ann. Phys. (N.Y.)* **2019**, *411*, 167998.

(4) Okuta, R.; Unno, Y.; Nishino, D.; Hido, S.; Loomis, C. CuPy: A NumPy-Compatible Library for NVIDIA GPU Calculations. Proceedings of Workshop on Machine Learning Systems (LearningSys) in The Thirty-first Annual Conference on Neural Information Processing Systems (NIPS). 2017; Code available at https://github.com/cupy/cupy.

(5) Shalashilin, D. V.; Child, M. S. Time Dependent Quantum Propagation in Phase Space. *J. Chem. Phys.* **2000**, *113*, 10028–10036.

(6) Shalashilin, D. V.; Child, M. S. Real Time Quantum Propagation on a Monte Carlo Trajectory Guided Grids of Coupled Coherent States: 26D Simulation of Pyrazine Absorption Spectrum. *J. Chem. Phys.* **2004**, *121*, 3563–3568.

(7) Ben-Nun, M.; Quenneville, J.; Martínez, T. J. Ab Initio Multiple Spawning: Photochemistry from First Principles Quantum Molecular Dynamics. *J. Phys. Chem. A* **2000**, *104*, 5161–5175.

(8) Gu, B.; Garashchuk, S. Quantum Dynamics with Gaussian Bases Defined by the Quantum Trajectories. *J. Phys. Chem. A* **2016**, *120*, 3023–3031.

(9) Werther, M.; Großmann, F. Apoptosis of Moving Nonorthogonal Basis Functions in Many-Particle Quantum Dynamics. *Phys. Rev. B* **2020**, *101*, 174315.

(10) Koch, W.; Frankcombe, T. J. Basis Expansion Leaping: A New Method to Solve the Time-Dependent Schrödinger Equation for Molecular Quantum Dynamics. *Phys. Rev. Lett.* **2013**, *110*, 263202.

(11) Richings, G.; Polyak, I.; Spinlove, K.; Worth, G.; Burghardt, I.; Lasorne, B. Quantum Dynamics Simulations Using Gaussian Wavepackets: the vMCG Method. *Int. Rev. Phys. Chem.* **2015**, *34*, 269–308.

(12) Saller, M. A. C.; Habershon, S. Quantum Dynamics with Short-Time Trajectories and Minimal Adaptive Basis Sets. *J. Chem. Theory Comput.* **2017**, *13*, 3085–3096.

(13) Hartke, B. Propagation with Distributed Gaussians as a Sparse, Adaptive Basis for Higher-Dimensional Quantum Dynamics. *Phys. Chem. Chem. Phys.* **2006**, *8*, 3627–3635.

(14) Sielk, J.; von Horsten, H. F.; Krüger, F.; Schneider, R.; Hartke, B. Quantum-Mechanical Wavepacket Propagation in a Sparse, Adaptive Basis of Interpolating Gaussians with Collocation. *Phys. Chem. Chem. Phys.* **2009**, *11*, 463–475.

(15) Dorfner, F.; Vidmar, L.; Brockt, C.; Jeckelmann, E.; Heidrich-Meisner, F. Real-Time Decay of a Highly Excited Charge Carrier in the One-Dimensional Holstein Model. *Phys. Rev. B* **2015**, *91*, 104302.

(16) Bonča, J.; Trugman, S. A.; Batistić, I. Holstein Polaron. *Phys. Rev. B* **1999**, *60*, 1633–1642.



\end{document}

%% file: main_contents.tex
The quantum dynamics of charge carriers and excitations in molecular systems is of great significance to a variety of research areas ranging from
physics to material science, chemistry and biology. 
Molecular vibrational degrees of freedom, which are quantized into phonons, can strongly influence such dynamics.
The Holstein model is a prototypical model for such electron--phonon (vibronic) coupling and
has been widely applied to various systems including polymers, molecular aggregates, and semiconductors~\cite{Coropceanu2007,Cheng2008,Ortmann2009,Schroeter2015,Huang2017,Hestand2018,Kundu2021,Li2021}.
A further crucial ingredient to such dynamics is an energy gradient or \enquote{tilt} representing,
for example, a voltage gradient in measurements of charge mobility or a natural energy funnel~\cite{Cao2020,cao106}.
Such biased vibronic systems exhibit various important coherent electron--phonon transport effects which have recently attracted much attention~\cite{Roulleau2011}, 
e.g., \enquote{phonon-assisted resonant tunneling} in inelastic tunneling experiments~\cite{Cui2015,Jung2015,Vdovin2016} or the \enquote{Franck--Condon (FC) blockade} in quantum dot or molecular tunneling experiments~\cite{Koch2005,Leturcq2009}, as well as vibrational enhancement of transport in antenna protein complexes~\cite{Kolli2012,Tiwari2013,Dijkstra,Cao2020}.
However, theoretical treatments of these phenomena have used perturbative approaches to calculate steady-state currents through a single site~\cite{Koch2005}, were restricted to single-phonon transitions~\cite{Emin1987} or were limited to the linear-response regime~\cite{cao150}.

In this paper, we report striking higher-order \emph{fractional} and \emph{long-range} resonances that occur within the \enquote{forbidden} FC blockaded regime when multiple sites are concatenated and their full non-equilibrium dynamics calculated non-perturbatively.
Previously, phononless long-range resonant tunneling has been studied theoretically and experimentally
in cold-gas quantum simulators on tilted optical lattices~\cite{Sachdev2002,Sias2007,Pielawa2011,Simon2011,Rubbo2011,Carrasquilla2013,Meinert2013,Meinert2014,Buyskikh2019,Buyskikh2019_2},
and such setups have been employed to investigate a plethora of phenomena such as quantum magnetism~\cite{Sachdev2002,Simon2011,Buyskikh2019,Buyskikh2019_2}, quantum dimer models~\cite{Pielawa2011}, transport properties and dynamical phase transition points~\cite{Gorshkov2011,Rubbo2011,Carrasquilla2013} or the creation of anyons~\cite{Keilmann2011}. 
We show that vibronic coupling naturally realizes and generalizes such resonant tunneling behavior.
\begin{figure}[t!]
\centering
\includegraphics
[scale=0.9]
{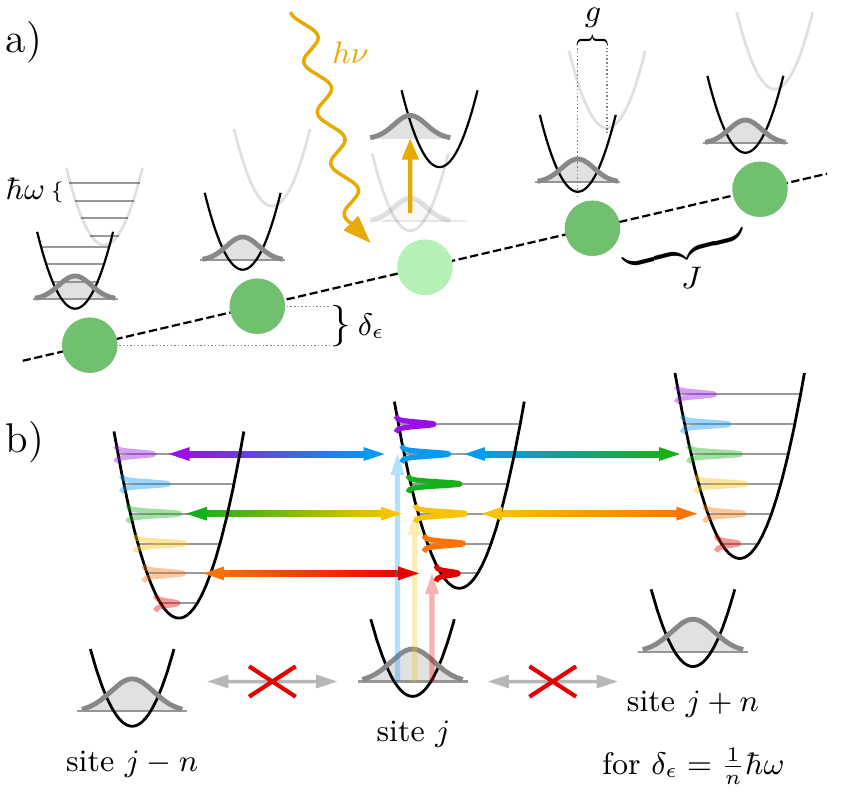}
\caption{a)~Sketch of the system: On the central site, a Franck--Condon (FC) excitation is induced, which tunnels to neighboring sites with amplitude $J$ and interacts with local phonons $\hbar \omega$ via Holstein coupling $g$ in the presence of a linear potential bias $\delta_\epsilon$.
b)~Resonant transitions are possible between local eigenstates of the excited PES. 
The FC excitation is a superposition of many different local eigenstates, allowing for multiple resonant vibronic transitions to both sides.
}
\label{fig:system}
\end{figure}
%
\paragraph{Model}\sectionstart%
Figure~\ref{fig:system}a) illustrates our setup: a chain of $L$ linearly tilted molecular sites which interact via nearest-neighbor coupling $J$ and couple linearly to local vibrations.
These vibrational degrees of freedom are represented by a single dominant mode, assumed to be harmonic with frequency $\omega$,
and each molecular site is coupled strongly to its own vibrational mode.
The coupling to the vibrations is quantified by the parameter $g$, related to the Huang--Rhys parameter as $S = \left(\frac{g}{\hbar\omega}\right)^2$.
A linear tilt shifts the energy of each site $j$ relative to site $j-1$ by $\delta_\epsilon$.
The resulting Hamiltonian is:
\begin{multline}
H = \sum_{j=0}^{L-1} \Bigg[ J \Big( \dyad{j}{j+1} + \text{h.c.} \Big) + \hbar\omega \left(b_j^\dag b_j + \frac{1}{2}\right) \\
    + g n^\text{exc}_j \left(b_j + b_j^\dag\right) + \delta_\epsilon\, j n_j^\text{exc} \Bigg]\, , \label{eq:Hamiltonian}
\end{multline}
where $\ket{j}$ is the excitonic state localized on site $j$ (where $\braket{i}{j} = \delta_{ij}$); $n^\text{exc}_j = \dyad{j}$ is the number of excitons on this site (restricted to 0 or 1); $b_j$ ($b_j^{\dag}$) are bosonic annihilation (creation) operators for phonons at site $j$; and $\hbar\omega \left(b_j^\dag b_j + \frac{1}{2}\right)$ is the local harmonic vibrational Hamiltonian on site $j$.
To accommodate the strong vibronic coupling, we truncate the vibrational Hilbert spaces at 128 phonons per mode, following ref.~\citenum{Kloss}.
We investigate the dynamics of a single Franck--Condon (vertical) excitation from the vibrational and electronic ground state,
which is initially (time $t=0$) localized to the central site.

The model allows for multiple interpretations in different contexts: 
$\ket{j}$ can represent the first excited state at the $j$-th molecule with electronic ground states on the remaining molecules.
However, $\ket{j}$ can also stand for a charged state arising from injecting an electron into the $j$-th molecular site, or for an atom placed at a certain position in an optical lattice, 
lending the findings possible relevance in various situations such as light harvesting, cold gas optical lattices, organic semiconductors, or non-equilibrium molecular junctions.

\paragraph{Numerical method}\sectionstart%
To present a complete dynamical picture, we have developed both a numerical algorithm and an analytical approach. 
Our massively parallelized numerical method is based on dynamically adapted effective basis sets, which can be grouped to the expanding family of adaptive basis methods (cf.~\cite{Shalashilin2000,Shalashilin2006,Ben-Nun,Gu2016,Werther2020,Koch2013,Richings,Saller2017,Hartke2006,Sielk2009}, see SI for details) but is technically most similar to a repeated dynamic use of \enquote{limited functional spaces}~\cite{Bonca,FHM2015}, and which can be applied in principle to arbitrary quantum dynamics problems.
By exploiting Hilbert-space localization of physically relevant states, the method computes dynamics only in the most relevant subspace, which is adaptively reconfigured to follow the evolving wavefunction.
The parallel nature of the linear-algebra problem is further leveraged by optimizing the method to be run on graphics processing units (GPUs), granting a substantial boost in runtime efficiency over existing methods, while still allowing treatment of both large local dimensions (up to $\nu_\text{max} = 255$) or large chain lengths (up to $L = 301$).
See the supporting information (SI) for further details.
To verify several important features of these numerical findings, we develop a random walk model with transition probabilities obtained from perturbation theory and evaluated using path integrals, as described below.

\paragraph{Resonance-enhanced transport and multimodal states}\sectionstart%
\begin{figure}
\centering
\includegraphics{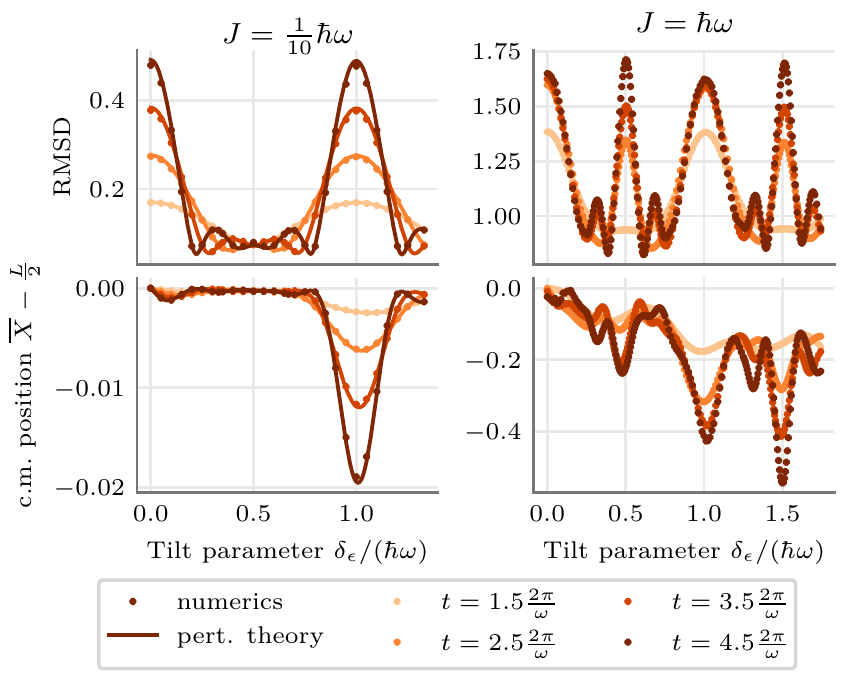}
\caption{The excitonic RMSD and movement of the average exciton position in a tilted Holstein chain for a series of tilt parameter values $\delta_\epsilon$ after different times $t$.
Left: perturbation theory (lines) and numerics (dots) for weak hopping $J = -\frac{1}{10} \hbar\omega$. Right: numerical data for strong hopping $J = -\hbar\omega$.
For all data: $L=9$, $g=4\hbar\omega$ (strong vibronic coupling).
}
\label{fig:perturbation}
\end{figure}%
We first calculate the Holstein dynamics for a range of values of the tilt parameter $\delta_\epsilon$ and two different (J-aggregate-like) values of $J$.
In Fig.~\ref{fig:perturbation}, the average position of the particle, $\overline{X} \coloneqq \sum_{j=0}^{L-1} j \expval{n^\text{exc}_j}$, and its root-mean-square deviation (RMSD),
\begin{equation*}
\text{RMSD} = \sqrt{\sigma^2} \coloneqq \sqrt{\sum_{j=0}^{L-1} \expval{n^\text{exc}_j} \left[j - \overline{X} \right]^2 } \,, 
\end{equation*}
are plotted as a function of the tilt parameter $\delta_\epsilon$ at different times.
Interestingly, the RMSD as a function of $\delta_\epsilon$ is highly non-linear, exhibiting strong spikes around integer and half-integer values of $\delta_\epsilon/(\hbar\omega)$. Additionally, smaller spikes appear around $\delta_\epsilon/(\hbar\omega) \in \{\frac{1}{3}, \frac{2}{3}, \frac{4}{3}\}$. Between these values, propagation is strongly suppressed, 
 suggesting transport-enhancing resonances are realized at certain values of 
 $\delta_\epsilon$.
Varying the remaining Hamiltonian parameters 
reveals that the location of the maxima depends only on the ratio $\delta_\epsilon/(\hbar\omega)$ 
(see SI section~2).
Note that the observed resonances for $\abs{\delta_\epsilon} < \hbar\omega$ lie within the \enquote{forbidden} FC blockade regime~\cite{Koch2005}.

\begin{figure*}
\centering
\includegraphics{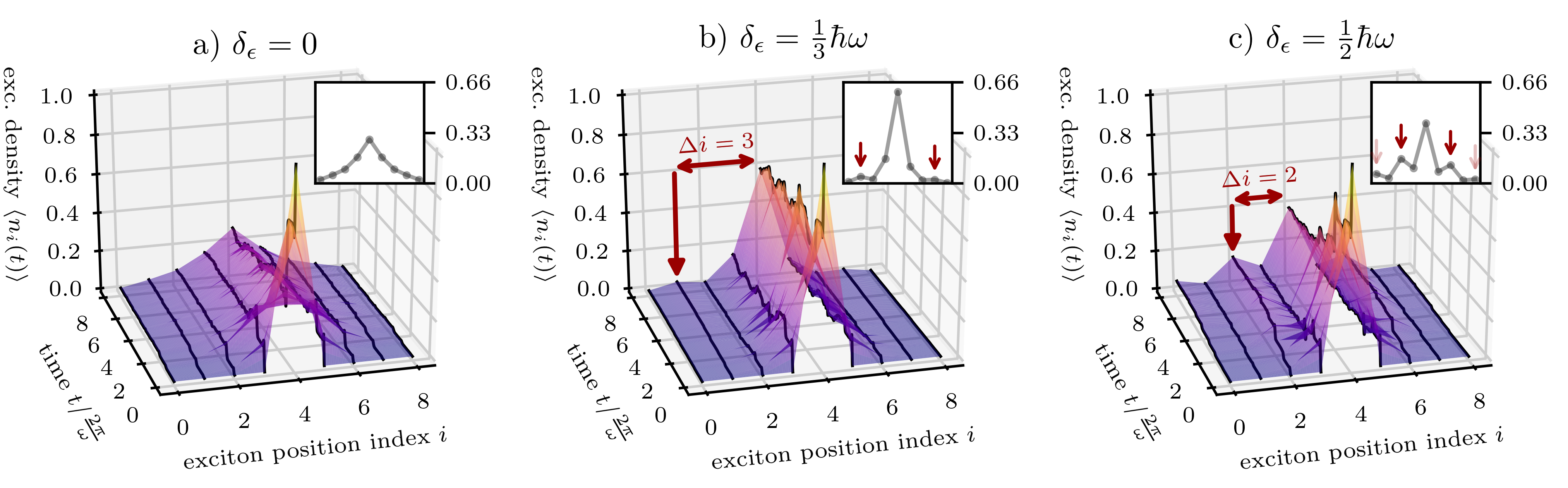}
\caption{The spatial exciton distribution over time under $J=\hbar\omega$ (as shown in Fig.~\ref{fig:perturbation}, right) for the a) homogeneous ($\delta_\epsilon = 0$) case, and the b) third-order resonant ($\delta_\epsilon = \frac{1}{3}\hbar\omega$) and c) second-order resonant ($\delta_\epsilon = \frac{1}{2}\hbar\omega$) cases. The insets show the exciton distribution at the final time $t = 8 \frac{2\pi}{\omega}$. In the second- and third-order resonant tilted cases, the exciton does not simply propagate outwards from a central peak into smooth tails, but instead excitonic density peaks form at the resonant sites, located $n$ sites from the initial site for $\delta_\epsilon = \frac{1}{n}\hbar\omega$.}
\label{fig:two_tilts}
\end{figure*}%
Furthermore, Fig.~\ref{fig:two_tilts} shows that the dynamics of the local density in the tilted chains behaves completely differently than in an untilted chain. 
For $\delta_\epsilon = 0$, the exciton propagates into a symmetric, peaked state that extends over a few lattice states, as shown in Fig.~\ref{fig:two_tilts}a) (cf.~\citenum{Kloss}). 
However, if $\delta_\epsilon$ is set to a resonant value, the final state 
is no longer a single-peak state, but exhibits multiple peaks and dips in its local exciton distribution. 
Further numerical results (see SI section 3) explicitly
show that the density $n$ sites off-center is maximized when $\delta_\epsilon$ is a multiple of $\frac{1}{n}\hbar\omega$. 

This brings us to our main insight: When the parameters are resonant as
$\delta_\epsilon/(\hbar\omega) = \frac{m}{n}$
with $n$ and $m$ small~\footnote{Based on our mechanistic explanation, we can expect to find the same resonance behavior and multimodal peaked states for any arbitrarily large integers $m$ and $n$ with $\frac{m}{n} = \delta_\epsilon/(\hbar\omega)$. 
    However, the necessary hopping strength $\abs{J}$ and the required system size to observe the resonance effects increase with $n$; thus, for finite system sizes and finite $J$, only small values of $n$ lead to observable resonance behavior. 
    Furthermore, the larger $m$, the smaller the number of resonant $\ket{\nu'}$ states in the uphill direction, leading to more pronounced asymmetry in the tunneling.} 
integers, then an excitation that is initially localized at site $j$ will 
tunnel to sites $j\pm n$ (and from there to sites $j\pm 2n$, etc.). 
In other words, 
the nearest-neighbor tunneling behavior of an unbiased chain 
is replaced by phonon-mediated (vibronic) tunneling over $n$ sites, leading to spatially structured, multimodal states---or effectively suppressing the tunneling if $n$ is too large or $\delta_\epsilon/(\hbar\omega) \notin \mathbb{Q}$. This $n$-site tunneling is the origin of the RMSD spikes seen in Fig.~\ref{fig:perturbation}.

Remarkably, certain resonant shifts $\delta_\epsilon$ increase the overall diffusivity, 
e.g., $\delta_\epsilon = \frac{\hbar\omega}{2}$ in 
Fig.~\ref{fig:perturbation}.
Propagation is then enhanced in both the \enquote{uphill} \emph{and} \enquote{downhill} direction, even for iterated resonant tunneling processes ($j \to j \pm 2 \to j \pm 4$), 
which might be exploited experimentally to engineer vibronically optimized transport.

\paragraph{The resonance mechanism}\sectionstart%
The observed resonant tunneling is due to matching energies between vibronic states at different sites, as illustrated in Fig.~\ref{fig:system}. 
Denoting the excited-state PES vibrational eigenstates as $\ket{\nu'}$ and those of the ground state PES as $\ket{\nu}$,
we find that the Franck--Condon $\ket{\nu = 0}$ state is a $g$-dependent superposition of many different $\ket{\nu'}$ eigenstates, given by the coherent state formula~\cite{Glauber}
$\ket{\nu = 0} \propto \sum_{\nu'} \frac{\alpha^{\nu'}}{\sqrt{\nu'!}} \ket{\nu'}$
with $\alpha = \frac{g}{\hbar \omega}$.
Then, if $\delta_\epsilon = \hbar\omega$, every constituent $\ket{\nu'}$ state of a Franck--Condon state is isoenergetic with the state $\ket{\nu'+1}$ at the downhill neighboring site and $\ket{\nu'-1}$ at the uphill neighbor, opening up multiple resonant transition pathways.
More generally, for $\delta_\epsilon = \frac{m}{n}\hbar \omega$, the resonant transitions are mediated by $m$-phonon and $n$-fold hopping matrix elements. 
Further, once the particle has tunneled from a state $\ket{\nu'}$ at site $j$ to $\ket{\nu'\pm m}$ at site $j \pm n$, the process can be successively repeated, tunneling to sites $\ket{\nu'\pm 2m}$ at $j \pm 2n$, etc. 

\begin{figure}
\centering
\includegraphics{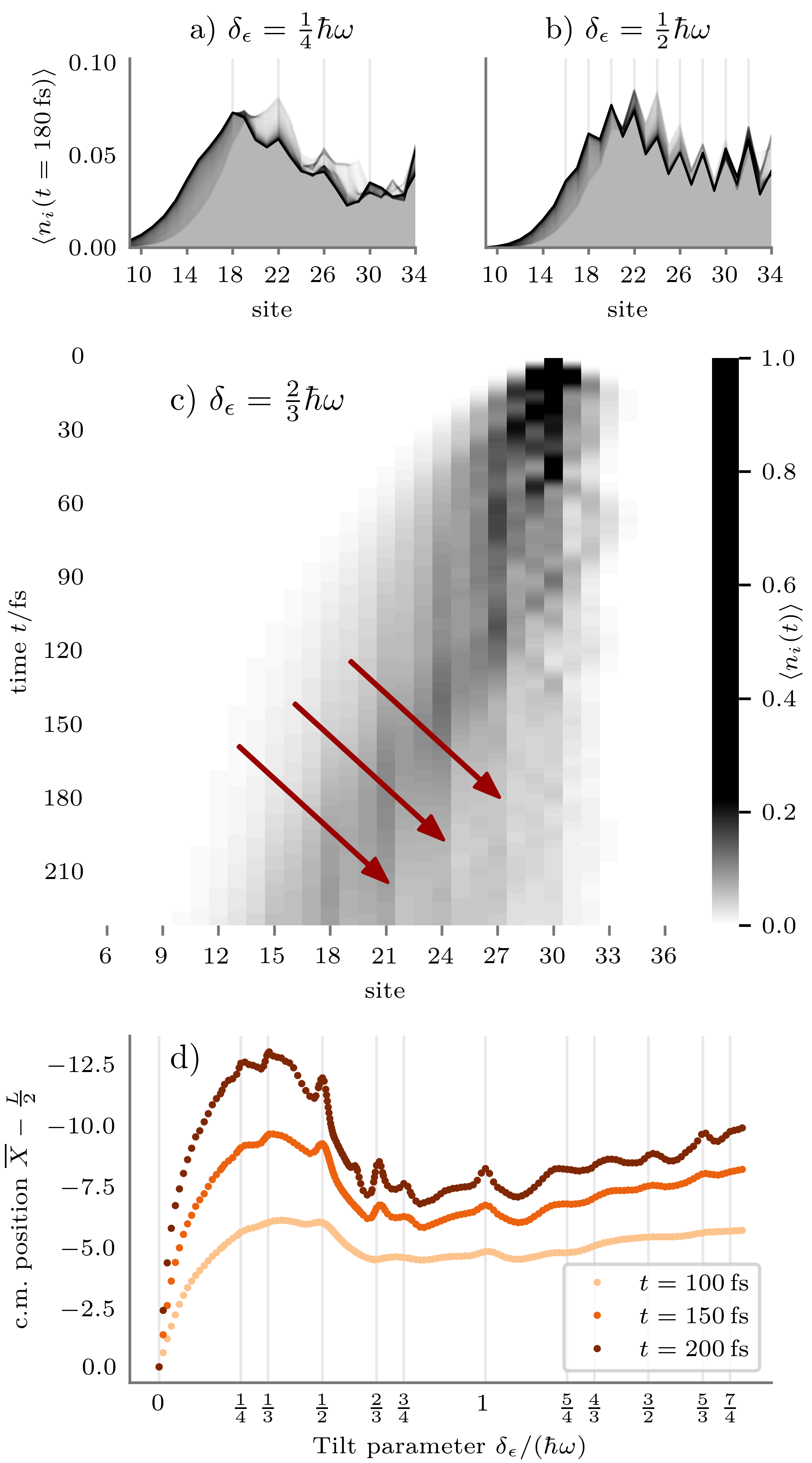}
\caption{Resonant tunneling with parameters corresponding to the DNA-scaffolded Cy3 system demonstrated by Hart et al.~\cite{Hart2021}. a), b)~The local exciton distribution \SI{180}{\fs} after initial excitation for two resonant energy gradients, with vertical lines as visual aids for the resonance interval, and shadows indicating the distribution over the preceding \SI{30}{\fs}. c)~The time evolution of the local exciton density for a 3-bond resonant tilt. Edges of resonance-induced high density are marked by arrows. d) Center-of-mass position for a range of tilts (cf.~Fig.~\ref{fig:perturbation} bottom) with simple rationals $\delta_\epsilon = \frac{m}{n}\hbar\omega$ up to $n = 4$ marked by vertical lines. $n=5$ resonances are also partially discernible.
The initial state is a Franck--Condon excitation from the center site [c), d)] or the last site [a), b)] for clarity. Parts of the chain not shown in a), b), c) for clarity. $\hbar\omega = \SI{1150}{\per\cm}$; $J = 0.55\hbar\omega = \SI{632.5}{\per\cm}$; $g = 0.71\hbar\omega$ ($S = 0.50$); $L = 35$ in a), b), $L=61$ in c), d).
}
\label{fig:Cy3}
\end{figure}

We can compare this behavior to the dynamics of a tilted Bose--Hubbard chain in the Mott insulating phase $\abs{J} \ll \abs{U}$ with equal filling $n_0$ at each site~\cite{Sachdev2002,Simon2011,Meinert2014}. 
When the tilt constant $\delta_\epsilon$ is a simple fraction of the Hubbard interaction, $\delta_\epsilon = U/n$, 
a single boson can tunnel from any site in the downhill direction by $n$ sites to form a state that is isoenergetic with the initial state~\cite{Meinert2014,Buyskikh2019_2}.
The resonant states can be mapped to dipoles and can be used to construct effective spin or quantum dimer models~\cite{Sachdev2002,Pielawa2011,Buyskikh2019_2}.
The resonant tunneling we observe in the tilted Holstein model is similar to
that in tilted Bose-Hubbard systems,
but differs in some key aspects:
\begin{enumerate*}[label=(\roman*)]
\item The multiple excited $\ket{\nu'}$ states that constitute the Franck--Condon excitation allow for tunneling in both the downhill \emph{and} uphill direction, 
whereas the Mott insulating state only allows for downhill transitions.
\item Our localized initial state induces tunneling from a single site.
\item Most importantly, the equidistant spacing of the vibrational QHO levels means that a repeated tunneling process is possible, since each tunneling event 
is energetically equivalent to an iterated tunneling, 
giving rise to the secondary resonance peaks in Fig.~\ref{fig:two_tilts}c) and Fig.~\ref{fig:Cy3}b). 
In contrast, in the Hubbard model, a doublon is bound to the first resonant site~\cite{Sachdev2002}.
\end{enumerate*}

\paragraph{Random-walk rate kernel model}\sectionstart%
To further support our findings, we verify the numerical results using an analytic approach based on hopping-rate kernels, which does not invoke the Markov approximation as in previous studies~\cite{cao104,cao123} and therefore belongs to the growing number of non-Markovian methods for open quantum systems (e.g., the transfer tensor method~\cite{Cerrillo2014}).
The resulting kinetics is equivalent to a continuous-time random walk, i.e., a generalization of Poisson kinetics on networks~\cite{Shlesinger1974,cao089}.

The basis of the approach is to first consider a dimer, $L = 2$, and separate the total Hamiltonian $H$ (eq~\ref{eq:Hamiltonian}) into the hopping operator $T = J \Big(\dyad{0}{1} + \dyad{1}{0} \Big)$ and $H_0 \coloneqq H - T$. 
Then the initial state is taken to be localized to $\ket{0}$ and the time-dependent transition probability is calculated as $q(t, \delta_\epsilon) \coloneqq \mel{1}{U^{(1)}(t)}{0}$ using the first-order propagator $U^{(1)}(t)$.
After tracing out the phononic degrees of freedom and evaluating the propagators using path integrals, we obtain
\begin{multline}
q(t, \delta_\epsilon)= \abs{\frac{J}{\hbar}}^2 \int^t_0 \dd{\tau_1} \int^t_0 \dd{\tau_2} \exp\!\Bigg\{ 2S \left(e^{-i\omega(\tau_1 - \tau_2)} - 1\right) \\
+2iS \left[ \sin(\omega\tau_1) - \sin(\omega\tau_2)\right]
 - i\frac{\delta_\epsilon}{\hbar} (\tau_1 - \tau_2)\!\Bigg\}, \label{eq:q(t)}
\end{multline}
where $S = \left(\frac{g}{\hbar\omega}\right)^2$ is the Huang--Rhys factor. 
These transition probabilities are extended to a chain of length $L > 2$ by applying the dimer transition probability on each bond.

To explain the vibronic resonance, we examine the properties of $q(t, \delta_\epsilon)$ in eq~\ref{eq:q(t)}.
Define $F(\tau_1, \tau_2)$ as the integrand of $q\fleft(t, \delta_\epsilon\fright)$ for $\delta_\epsilon = 0$.
Then $F\fleft(\tau_1, \tau_2\fright)$ is periodic in both arguments with a period $T=\frac{2\pi}{\omega}$.
Now, for general $\delta_\epsilon \in \mathbb{R}$,
we can write the transition probability in eq~\ref{eq:q(t)}  as
the 2D Fourier transform of $\chi_{\left[0, t \right]^2} F(\tau_1, \tau_2)$ evaluated at $\left(\frac{\delta_\epsilon}{\hbar}, -\frac{\delta_\epsilon}{\hbar}\right)$,
where $\chi_{\left[0, t \right]^2}$ is the two-dimensional boxcar function on $\tau_1, \tau_2$. 
In the absence of $\chi_{\left[0, t \right]^2}$, the transfer probability $q(t, \delta_\epsilon)$ 
would vanish unless the Fourier frequency matches the periodicity of $F$:
\[ \frac{\delta_\epsilon}{\hbar} = m \frac{2\pi}{T } = m \omega \qfor m \in \mathbb{N}_{0}.\]
This is the first-order vibronic resonance condition and demonstrates how the resonance peaks arise. 
When taking into account $\chi_{\left[0, t \right]^2}$, which acts as a 2D convolution in frequency space,
and using the Dirac comb structure of $\hat{F}$, 
 we find that
\begin{equation}
\eval{q(t, \delta_\epsilon)}_{\delta_\epsilon \approx m \hbar \omega} \approx 2 c_m \frac{1 - \cos(t \left(\frac{\delta_\epsilon}{\hbar} - m \omega\right))}{\left(\frac{\delta_\epsilon}{\hbar} - m \omega\right)^2},\label{eq:q}
\end{equation}
where $c_m$ is the Fourier coefficient of $F$ at $\omega_1 = -\omega_2 = m\omega$.
Equation~\ref{eq:q} determines the structure of a resonance peak at $\delta_\epsilon = m\hbar\omega$ and shows the oscillatory structure of the transient side-peaks around the main resonance peaks. 
Furthermore, one can easily see from a Taylor expansion of the cosine that the main peaks at $\delta_\epsilon = m\hbar\omega$ become sharper and taller with $\sim t^2$.
Both the oscillatory side-peaks and the quadratic growth of the main peaks are confirmed in Fig.~\ref{fig:perturbation}a).

Physically, the first-order perturbation describes the transition between adjacent sites, which differ in energy by $\delta_\epsilon$. In accord with energy conservation, the transition is allowed if the vibrational energy difference $m\hbar\omega$ matches the tilt energy,  $m\hbar\omega=\delta_\epsilon$.
More generally, to capture long-range tunneling over $n$ bonds, an $n$-th order perturbative expansion would be required, leading to the generalized fractional resonance condition
$m\hbar\omega=n\delta_\epsilon$.

In Fig.~\ref{fig:perturbation}, we compare the perturbative dynamics to the numerical results for $J = -\frac{1}{10} \hbar\omega$, 
which shows excellent agreement between the results.
Both methods show first-order tunneling spikes around integer multiples of $\hbar\omega$. 
The weak $J$ suppresses higher-order tunneling events and the associated spikes at rational fractions $\frac{m}{n}\hbar\omega$
(though \emph{transient} side-peaks appear, predicted by eq~\ref{eq:q}).
A small but persistent second-order tunneling effect is observable only
in the full numerical calculations (see SI section~4).
For $\delta_\epsilon \lessapprox 0.1 \hbar\omega$, we observe linear response behavior for short times (SI section~5).

\paragraph{Application to engineered Cy3 system}\sectionstart%
The analysis so far has clearly demonstrated our mechanism for an abstract Holstein model.
To investigate its impact in realistic systems, we choose an oligomeric extension of the recently demonstrated DNA-scaffolded tunable Cy3 dyes~\cite{Hart2021,Hart2022} as a model system and numerically calculate the dynamics using the adjacent-dimer (0-nt) parameters for each nearest-neighbor interaction: $\hbar\omega = \SI{1150}{\per\cm}$, $J = 0.55\hbar\omega = \SI{632.5}{\per\cm}$ (H-aggregate coupling) and $g = 0.71\hbar\omega$ ($S = 0.50$).
Since the polaronic trapping is much weaker for this smaller Huang--Rhys factor $S$, the chain length is increased from $L=9$ up to $L=61$ to eliminate boundary effects while $\nu_\text{max}$ is decreased to 16.
The results of these calculations can be seen in Fig.~\ref{fig:Cy3}.
We see the same resonance features at simple rational fractions $\delta_\epsilon = \frac{m}{n} \hbar\omega$, including more long-ranged (higher $n$) resonances enabled by the greater chain length.
The features are clearly visible albeit less distinct than in Fig.~\ref{fig:perturbation}, as is to be expected due to the decreased dominance of the vibronic dynamics over the excitonic dynamics.
This also causes Bloch~\cite{Nieuwenburg2019} (or Wannier--Stark~\cite{Bleuse1988,Mendez1988,Weiser1992}) localization to arise, which is the reason behind the momentary \emph{decrease} in transport as $\delta_\epsilon$ increases beyond $\frac{\hbar\omega}{3}$.
These results provide a proof-of-principle for a typical and realistic parameter regime, but we note that our model assumes fully coherent transport and does not include disorder or further bath modes, which would be the subject of future analysis and may impact the results in this particular hypothetical model system.

\paragraph{Conclusion \& Outlook}\sectionstart%
We have uncovered resonance-dependent transport behavior in tilted vibronic chains. Tunneling over $n$ bonds is allowed for $\delta_\epsilon/(\hbar\omega) = \frac{m}{n}$, corresponding to $m$-phonon and $n$-th order tunneling transitions. 
To study this problem, we have developed both an analytical and a numerical method. 

This generalizes the resonant tunneling found in Mott insulators on tilted Bose--Hubbard chains~\cite{Sachdev2002,Simon2011,Meinert2014}, as long-range, repeated hopping in both directions is naturally obtained.
Vibronic coherence has also emerged as an active mechanism in light-harvesting systems, molecular semiconductors, and molecular electronics.  
Our discovery of long-range tunneling resonances have an important bearing on the \enquote{phonon antenna} mechanism~\cite{Chin2012,delRey2013}, a new type of environment-assisted quantum transport~\cite{QuEB}.
Prospective technological applications are to exploit the bias-dependent resonance peaks for optimization or selective switching of quantum transport, or to enable nanoscale sensing of structural parameters, e.g., as an extension of inelastic electron tunneling spectroscopy (cf.~\citenum{Galperin2007}).
Future efforts will aim at extensions from the chain configuration to thin films, nanotubes, and quantum networks, and determine the influence of coupling to further bath modes (cf.~\citenum{Chuang2016}): 
Calculations of noisy driven energy transfer in a dimer suggest that such behavior is indeed robust~\cite{Dijkstra}.
Furthermore, we expect this behavior can be realized in quantum simulators using tilted optical lattices~\cite{Pazy2005,Schaefer2020,Herrera2011} or superconducting qubits~\cite{Mostame2012,Potocnik2018,Wang2020}.

%% file: ms.bbl
%

%% file: ms.bbl
\begin{thebibliography}{70}%
\makeatletter
\providecommand \@ifxundefined [1]{%
 \@ifx{#1\undefined}
}%
\providecommand \@ifnum [1]{%
 \ifnum #1\expandafter \@firstoftwo
 \else \expandafter \@secondoftwo
 \fi
}%
\providecommand \@ifx [1]{%
 \ifx #1\expandafter \@firstoftwo
 \else \expandafter \@secondoftwo
 \fi
}%
\providecommand \natexlab [1]{#1}%
\providecommand \enquote  [1]{``#1''}%
\providecommand \bibnamefont  [1]{#1}%
\providecommand \bibfnamefont [1]{#1}%
\providecommand \citenamefont [1]{#1}%
\providecommand \href@noop [0]{\@secondoftwo}%
\providecommand \href [0]{\begingroup \@sanitize@url \@href}%
\providecommand \@href[1]{\@@startlink{#1}\@@href}%
\providecommand \@@href[1]{\endgroup#1\@@endlink}%
\providecommand \@sanitize@url [0]{\catcode `\\12\catcode `\$12\catcode
  `\&12\catcode `\#12\catcode `\^12\catcode `\_12\catcode `\%12\relax}%
\providecommand \@@startlink[1]{}%
\providecommand \@@endlink[0]{}%
\providecommand \url  [0]{\begingroup\@sanitize@url \@url }%
\providecommand \@url [1]{\endgroup\@href {#1}{\urlprefix }}%
\providecommand \urlprefix  [0]{URL }%
\providecommand \Eprint [0]{\href }%
\providecommand \doibase [0]{https://doi.org/}%
\providecommand \selectlanguage [0]{\@gobble}%
\providecommand \bibinfo  [0]{\@secondoftwo}%
\providecommand \bibfield  [0]{\@secondoftwo}%
\providecommand \translation [1]{[#1]}%
\providecommand \BibitemOpen [0]{}%
\providecommand \bibitemStop [0]{}%
\providecommand \bibitemNoStop [0]{.\EOS\space}%
\providecommand \EOS [0]{\spacefactor3000\relax}%
\providecommand \BibitemShut  [1]{\csname bibitem#1\endcsname}%
\let\auto@bib@innerbib\@empty
\bibitem [{\citenamefont {Coropceanu}\ \emph {et~al.}(2007)\citenamefont
  {Coropceanu}, \citenamefont {Cornil}, \citenamefont {da~Silva~Filho},
  \citenamefont {Olivier}, \citenamefont {Silbey},\ and\ \citenamefont
  {Brédas}}]{Coropceanu2007}%
  \BibitemOpen
  \bibfield  {author} {\bibinfo {author} {\bibfnamefont {V.}~\bibnamefont
  {Coropceanu}}, \bibinfo {author} {\bibfnamefont {J.}~\bibnamefont {Cornil}},
  \bibinfo {author} {\bibfnamefont {D.~A.}\ \bibnamefont {da~Silva~Filho}},
  \bibinfo {author} {\bibfnamefont {Y.}~\bibnamefont {Olivier}}, \bibinfo
  {author} {\bibfnamefont {R.}~\bibnamefont {Silbey}},\ and\ \bibinfo {author}
  {\bibfnamefont {J.-L.}\ \bibnamefont {Brédas}},\ }\bibfield  {title}
  {\bibinfo {title} {Charge transport in organic semiconductors},\ }\href
  {https://doi.org/10.1021/cr050140x} {\bibfield  {journal} {\bibinfo
  {journal} {Chem.\ Rev.}\ }\textbf {\bibinfo {volume} {107}},\ \bibinfo
  {pages} {926} (\bibinfo {year} {2007})}\BibitemShut {NoStop}%
\bibitem [{\citenamefont {Cheng}\ and\ \citenamefont
  {Silbey}(2008)}]{Cheng2008}%
  \BibitemOpen
  \bibfield  {author} {\bibinfo {author} {\bibfnamefont {Y.-C.}\ \bibnamefont
  {Cheng}}\ and\ \bibinfo {author} {\bibfnamefont {R.~J.}\ \bibnamefont
  {Silbey}},\ }\bibfield  {title} {\bibinfo {title} {A unified theory for
  charge-carrier transport in organic crystals},\ }\href
  {https://doi.org/10.1063/1.2894840} {\bibfield  {journal} {\bibinfo
  {journal} {J.\ Chem.\ Phys.}\ }\textbf {\bibinfo {volume} {128}},\ \bibinfo
  {pages} {114713} (\bibinfo {year} {2008})}\BibitemShut {NoStop}%
\bibitem [{\citenamefont {Ortmann}\ \emph {et~al.}(2009)\citenamefont
  {Ortmann}, \citenamefont {Bechstedt},\ and\ \citenamefont
  {Hannewald}}]{Ortmann2009}%
  \BibitemOpen
  \bibfield  {author} {\bibinfo {author} {\bibfnamefont {F.}~\bibnamefont
  {Ortmann}}, \bibinfo {author} {\bibfnamefont {F.}~\bibnamefont {Bechstedt}},\
  and\ \bibinfo {author} {\bibfnamefont {K.}~\bibnamefont {Hannewald}},\
  }\bibfield  {title} {\bibinfo {title} {Theory of charge transport in organic
  crystals: Beyond holstein's small-polaron model},\ }\href
  {https://doi.org/10.1103/PhysRevB.79.235206} {\bibfield  {journal} {\bibinfo
  {journal} {Phys.\ Rev.\ B}\ }\textbf {\bibinfo {volume} {79}},\ \bibinfo
  {pages} {235206} (\bibinfo {year} {2009})}\BibitemShut {NoStop}%
\bibitem [{\citenamefont {Schröter}\ \emph {et~al.}(2015)\citenamefont
  {Schröter}, \citenamefont {Ivanov}, \citenamefont {Schulze}, \citenamefont
  {Polyutov}, \citenamefont {Yan}, \citenamefont {Pullerits},\ and\
  \citenamefont {Kühn}}]{Schroeter2015}%
  \BibitemOpen
  \bibfield  {author} {\bibinfo {author} {\bibfnamefont {M.}~\bibnamefont
  {Schröter}}, \bibinfo {author} {\bibfnamefont {S.}~\bibnamefont {Ivanov}},
  \bibinfo {author} {\bibfnamefont {J.}~\bibnamefont {Schulze}}, \bibinfo
  {author} {\bibfnamefont {S.}~\bibnamefont {Polyutov}}, \bibinfo {author}
  {\bibfnamefont {Y.}~\bibnamefont {Yan}}, \bibinfo {author} {\bibfnamefont
  {T.}~\bibnamefont {Pullerits}},\ and\ \bibinfo {author} {\bibfnamefont
  {O.}~\bibnamefont {Kühn}},\ }\bibfield  {title} {\bibinfo {title}
  {Exciton--vibrational coupling in the dynamics and spectroscopy of frenkel
  excitons in molecular aggregates},\ }\href
  {https://doi.org/https://doi.org/10.1016/j.physrep.2014.12.001} {\bibfield
  {journal} {\bibinfo  {journal} {Phys.\ Rep.}\ }\textbf {\bibinfo {volume}
  {567}},\ \bibinfo {pages} {1} (\bibinfo {year} {2015})}\BibitemShut {NoStop}%
\bibitem [{\citenamefont {Huang}\ \emph {et~al.}(2017)\citenamefont {Huang},
  \citenamefont {Chen}, \citenamefont {Zhou},\ and\ \citenamefont
  {Zhao}}]{Huang2017}%
  \BibitemOpen
  \bibfield  {author} {\bibinfo {author} {\bibfnamefont {Z.}~\bibnamefont
  {Huang}}, \bibinfo {author} {\bibfnamefont {L.}~\bibnamefont {Chen}},
  \bibinfo {author} {\bibfnamefont {N.}~\bibnamefont {Zhou}},\ and\ \bibinfo
  {author} {\bibfnamefont {Y.}~\bibnamefont {Zhao}},\ }\bibfield  {title}
  {\bibinfo {title} {Transient dynamics of a one-dimensional holstein polaron
  under the influence of an external electric field},\ }\href
  {https://doi.org/10.1002/andp.201600367} {\bibfield  {journal} {\bibinfo
  {journal} {Ann.\ Phys.\ (Berlin)}\ }\textbf {\bibinfo {volume} {529}},\
  \bibinfo {pages} {1600367} (\bibinfo {year} {2017})}\BibitemShut {NoStop}%
\bibitem [{\citenamefont {Hestand}\ and\ \citenamefont
  {Spano}(2018)}]{Hestand2018}%
  \BibitemOpen
  \bibfield  {author} {\bibinfo {author} {\bibfnamefont {N.~J.}\ \bibnamefont
  {Hestand}}\ and\ \bibinfo {author} {\bibfnamefont {F.~C.}\ \bibnamefont
  {Spano}},\ }\bibfield  {title} {\bibinfo {title} {Expanded theory of h- and
  j-molecular aggregates: The effects of vibronic coupling and intermolecular
  charge transfer},\ }\href {https://doi.org/10.1021/acs.chemrev.7b00581}
  {\bibfield  {journal} {\bibinfo  {journal} {Chem.\ Rev.}\ }\textbf {\bibinfo
  {volume} {118}},\ \bibinfo {pages} {7069} (\bibinfo {year}
  {2018})}\BibitemShut {NoStop}%
\bibitem [{\citenamefont {Kundu}\ and\ \citenamefont
  {Makri}(2021)}]{Kundu2021}%
  \BibitemOpen
  \bibfield  {author} {\bibinfo {author} {\bibfnamefont {S.}~\bibnamefont
  {Kundu}}\ and\ \bibinfo {author} {\bibfnamefont {N.}~\bibnamefont {Makri}},\
  }\bibfield  {title} {\bibinfo {title} {Exciton--vibration dynamics in
  j-aggregates of a perylene bisimide from real-time path integral
  calculations},\ }\href {https://doi.org/10.1021/acs.jpcc.0c09405} {\bibfield
  {journal} {\bibinfo  {journal} {J.\ Phys.\ Chem.\ C}\ }\textbf {\bibinfo
  {volume} {125}},\ \bibinfo {pages} {201} (\bibinfo {year}
  {2021})}\BibitemShut {NoStop}%
\bibitem [{\citenamefont {Li}\ \emph {et~al.}(2021)\citenamefont {Li},
  \citenamefont {Ren},\ and\ \citenamefont {Shuai}}]{Li2021}%
  \BibitemOpen
  \bibfield  {author} {\bibinfo {author} {\bibfnamefont {W.}~\bibnamefont
  {Li}}, \bibinfo {author} {\bibfnamefont {J.}~\bibnamefont {Ren}},\ and\
  \bibinfo {author} {\bibfnamefont {Z.}~\bibnamefont {Shuai}},\ }\bibfield
  {title} {\bibinfo {title} {A general charge transport picture for organic
  semiconductors with nonlocal electron--phonon couplings},\ }\href
  {https://doi.org/10.1038/s41467-021-24520-y} {\bibfield  {journal} {\bibinfo
  {journal} {Nat.\ Commun.}\ }\textbf {\bibinfo {volume} {12}},\ \bibinfo
  {pages} {4260} (\bibinfo {year} {2021})}\BibitemShut {NoStop}%
\bibitem [{\citenamefont {Cao}\ \emph {et~al.}(2020)\citenamefont {Cao},
  \citenamefont {Cogdell}, \citenamefont {Coker}, \citenamefont {Duan},
  \citenamefont {Hauer}, \citenamefont {Kleinekathöfer}, \citenamefont
  {Jansen}, \citenamefont {Mančal}, \citenamefont {Miller}, \citenamefont
  {Ogilvie}, \citenamefont {Prokhorenko}, \citenamefont {Renger}, \citenamefont
  {Tan}, \citenamefont {Tempelaar}, \citenamefont {Thorwart}, \citenamefont
  {Thyrhaug}, \citenamefont {Westenhoff},\ and\ \citenamefont
  {Zigmantas}}]{Cao2020}%
  \BibitemOpen
  \bibfield  {author} {\bibinfo {author} {\bibfnamefont {J.}~\bibnamefont
  {Cao}}, \bibinfo {author} {\bibfnamefont {R.~J.}\ \bibnamefont {Cogdell}},
  \bibinfo {author} {\bibfnamefont {D.~F.}\ \bibnamefont {Coker}}, \bibinfo
  {author} {\bibfnamefont {H.-G.}\ \bibnamefont {Duan}}, \bibinfo {author}
  {\bibfnamefont {J.}~\bibnamefont {Hauer}}, \bibinfo {author} {\bibfnamefont
  {U.}~\bibnamefont {Kleinekathöfer}}, \bibinfo {author} {\bibfnamefont
  {T.~L.~C.}\ \bibnamefont {Jansen}}, \bibinfo {author} {\bibfnamefont
  {T.}~\bibnamefont {Mančal}}, \bibinfo {author} {\bibfnamefont {R.~J.~D.}\
  \bibnamefont {Miller}}, \bibinfo {author} {\bibfnamefont {J.~P.}\
  \bibnamefont {Ogilvie}}, \bibinfo {author} {\bibfnamefont {V.~I.}\
  \bibnamefont {Prokhorenko}}, \bibinfo {author} {\bibfnamefont
  {T.}~\bibnamefont {Renger}}, \bibinfo {author} {\bibfnamefont {H.-S.}\
  \bibnamefont {Tan}}, \bibinfo {author} {\bibfnamefont {R.}~\bibnamefont
  {Tempelaar}}, \bibinfo {author} {\bibfnamefont {M.}~\bibnamefont {Thorwart}},
  \bibinfo {author} {\bibfnamefont {E.}~\bibnamefont {Thyrhaug}}, \bibinfo
  {author} {\bibfnamefont {S.}~\bibnamefont {Westenhoff}},\ and\ \bibinfo
  {author} {\bibfnamefont {D.}~\bibnamefont {Zigmantas}},\ }\bibfield  {title}
  {\bibinfo {title} {Quantum biology revisited},\ }\bibfield  {journal}
  {\bibinfo  {journal} {Sci.\ Adv.}\ }\textbf {\bibinfo {volume} {6}},\ \href
  {https://doi.org/10.1126/sciadv.aaz4888} {10.1126/sciadv.aaz4888} (\bibinfo
  {year} {2020})\BibitemShut {NoStop}%
\bibitem [{\citenamefont {Wu}\ \emph {et~al.}(2010)\citenamefont {Wu},
  \citenamefont {Liu}, \citenamefont {Shen}, \citenamefont {Cao},\ and\
  \citenamefont {Silbey}}]{cao106}%
  \BibitemOpen
  \bibfield  {author} {\bibinfo {author} {\bibfnamefont {J.}~\bibnamefont
  {Wu}}, \bibinfo {author} {\bibfnamefont {F.}~\bibnamefont {Liu}}, \bibinfo
  {author} {\bibfnamefont {Y.}~\bibnamefont {Shen}}, \bibinfo {author}
  {\bibfnamefont {J.}~\bibnamefont {Cao}},\ and\ \bibinfo {author}
  {\bibfnamefont {R.~J.}\ \bibnamefont {Silbey}},\ }\bibfield  {title}
  {\bibinfo {title} {Efficient energy transfer in light-harvesting systems, i:
  Optimal temperature, reorganization energy and spatial{\textendash}temporal
  correlations},\ }\href {https://doi.org/10.1088/1367-2630/12/10/105012}
  {\bibfield  {journal} {\bibinfo  {journal} {New J.\ Phys.}\ }\textbf
  {\bibinfo {volume} {12}},\ \bibinfo {pages} {105012} (\bibinfo {year}
  {2010})}\BibitemShut {NoStop}%
\bibitem [{\citenamefont {Roulleau}\ \emph {et~al.}(2011)\citenamefont
  {Roulleau}, \citenamefont {Baer}, \citenamefont {Choi}, \citenamefont
  {Molitor}, \citenamefont {G{\"u}ttinger}, \citenamefont {M{\"u}ller},
  \citenamefont {Dr{\"o}scher}, \citenamefont {Ensslin},\ and\ \citenamefont
  {Ihn}}]{Roulleau2011}%
  \BibitemOpen
  \bibfield  {author} {\bibinfo {author} {\bibfnamefont {P.}~\bibnamefont
  {Roulleau}}, \bibinfo {author} {\bibfnamefont {S.}~\bibnamefont {Baer}},
  \bibinfo {author} {\bibfnamefont {T.}~\bibnamefont {Choi}}, \bibinfo {author}
  {\bibfnamefont {F.}~\bibnamefont {Molitor}}, \bibinfo {author} {\bibfnamefont
  {J.}~\bibnamefont {G{\"u}ttinger}}, \bibinfo {author} {\bibfnamefont
  {T.}~\bibnamefont {M{\"u}ller}}, \bibinfo {author} {\bibfnamefont
  {S.}~\bibnamefont {Dr{\"o}scher}}, \bibinfo {author} {\bibfnamefont
  {K.}~\bibnamefont {Ensslin}},\ and\ \bibinfo {author} {\bibfnamefont
  {T.}~\bibnamefont {Ihn}},\ }\bibfield  {title} {\bibinfo {title} {Coherent
  electron--phonon coupling in tailored quantum systems},\ }\href
  {https://doi.org/10.1038/ncomms1241} {\bibfield  {journal} {\bibinfo
  {journal} {Nat.\ Commun.}\ }\textbf {\bibinfo {volume} {2}},\ \bibinfo
  {pages} {239} (\bibinfo {year} {2011})}\BibitemShut {NoStop}%
\bibitem [{\citenamefont {Cui}\ \emph {et~al.}(2015)\citenamefont {Cui},
  \citenamefont {Tosoni}, \citenamefont {Schneider}, \citenamefont {Pacchioni},
  \citenamefont {Nilius},\ and\ \citenamefont {Freund}}]{Cui2015}%
  \BibitemOpen
  \bibfield  {author} {\bibinfo {author} {\bibfnamefont {Y.}~\bibnamefont
  {Cui}}, \bibinfo {author} {\bibfnamefont {S.}~\bibnamefont {Tosoni}},
  \bibinfo {author} {\bibfnamefont {W.-D.}\ \bibnamefont {Schneider}}, \bibinfo
  {author} {\bibfnamefont {G.}~\bibnamefont {Pacchioni}}, \bibinfo {author}
  {\bibfnamefont {N.}~\bibnamefont {Nilius}},\ and\ \bibinfo {author}
  {\bibfnamefont {H.-J.}\ \bibnamefont {Freund}},\ }\bibfield  {title}
  {\bibinfo {title} {Phonon-mediated electron transport through cao thin
  films},\ }\href {https://doi.org/10.1103/PhysRevLett.114.016804} {\bibfield
  {journal} {\bibinfo  {journal} {Phys.\ Rev.\ Lett.}\ }\textbf {\bibinfo
  {volume} {114}},\ \bibinfo {pages} {016804} (\bibinfo {year}
  {2015})}\BibitemShut {NoStop}%
\bibitem [{\citenamefont {Jung}\ \emph {et~al.}(2015)\citenamefont {Jung},
  \citenamefont {Park}, \citenamefont {Park}, \citenamefont {Jeong},
  \citenamefont {Kim}, \citenamefont {Watanabe}, \citenamefont {Taniguchi},
  \citenamefont {Ha}, \citenamefont {Hwang},\ and\ \citenamefont
  {Kim}}]{Jung2015}%
  \BibitemOpen
  \bibfield  {author} {\bibinfo {author} {\bibfnamefont {S.}~\bibnamefont
  {Jung}}, \bibinfo {author} {\bibfnamefont {M.}~\bibnamefont {Park}}, \bibinfo
  {author} {\bibfnamefont {J.}~\bibnamefont {Park}}, \bibinfo {author}
  {\bibfnamefont {T.-Y.}\ \bibnamefont {Jeong}}, \bibinfo {author}
  {\bibfnamefont {H.-J.}\ \bibnamefont {Kim}}, \bibinfo {author} {\bibfnamefont
  {K.}~\bibnamefont {Watanabe}}, \bibinfo {author} {\bibfnamefont
  {T.}~\bibnamefont {Taniguchi}}, \bibinfo {author} {\bibfnamefont {D.~H.}\
  \bibnamefont {Ha}}, \bibinfo {author} {\bibfnamefont {C.}~\bibnamefont
  {Hwang}},\ and\ \bibinfo {author} {\bibfnamefont {Y.-S.}\ \bibnamefont
  {Kim}},\ }\bibfield  {title} {\bibinfo {title} {Vibrational properties of
  h-bn and h-bn-graphene heterostructures probed by inelastic electron
  tunneling spectroscopy},\ }\href {https://doi.org/10.1038/srep16642}
  {\bibfield  {journal} {\bibinfo  {journal} {Sci.\ Rep.}\ }\textbf {\bibinfo
  {volume} {5}},\ \bibinfo {pages} {16642} (\bibinfo {year}
  {2015})}\BibitemShut {NoStop}%
\bibitem [{\citenamefont {Vdovin}\ \emph {et~al.}(2016)\citenamefont {Vdovin},
  \citenamefont {Mishchenko}, \citenamefont {Greenaway}, \citenamefont {Zhu},
  \citenamefont {Ghazaryan}, \citenamefont {Misra}, \citenamefont {Cao},
  \citenamefont {Morozov}, \citenamefont {Makarovsky}, \citenamefont
  {Fromhold}, \citenamefont {Patan\`e}, \citenamefont {Slotman}, \citenamefont
  {Katsnelson}, \citenamefont {Geim}, \citenamefont {Novoselov},\ and\
  \citenamefont {Eaves}}]{Vdovin2016}%
  \BibitemOpen
  \bibfield  {author} {\bibinfo {author} {\bibfnamefont {E.~E.}\ \bibnamefont
  {Vdovin}}, \bibinfo {author} {\bibfnamefont {A.}~\bibnamefont {Mishchenko}},
  \bibinfo {author} {\bibfnamefont {M.~T.}\ \bibnamefont {Greenaway}}, \bibinfo
  {author} {\bibfnamefont {M.~J.}\ \bibnamefont {Zhu}}, \bibinfo {author}
  {\bibfnamefont {D.}~\bibnamefont {Ghazaryan}}, \bibinfo {author}
  {\bibfnamefont {A.}~\bibnamefont {Misra}}, \bibinfo {author} {\bibfnamefont
  {Y.}~\bibnamefont {Cao}}, \bibinfo {author} {\bibfnamefont {S.~V.}\
  \bibnamefont {Morozov}}, \bibinfo {author} {\bibfnamefont {O.}~\bibnamefont
  {Makarovsky}}, \bibinfo {author} {\bibfnamefont {T.~M.}\ \bibnamefont
  {Fromhold}}, \bibinfo {author} {\bibfnamefont {A.}~\bibnamefont {Patan\`e}},
  \bibinfo {author} {\bibfnamefont {G.~J.}\ \bibnamefont {Slotman}}, \bibinfo
  {author} {\bibfnamefont {M.~I.}\ \bibnamefont {Katsnelson}}, \bibinfo
  {author} {\bibfnamefont {A.~K.}\ \bibnamefont {Geim}}, \bibinfo {author}
  {\bibfnamefont {K.~S.}\ \bibnamefont {Novoselov}},\ and\ \bibinfo {author}
  {\bibfnamefont {L.}~\bibnamefont {Eaves}},\ }\bibfield  {title} {\bibinfo
  {title} {Phonon-assisted resonant tunneling of electrons in graphene--boron
  nitride transistors},\ }\href
  {https://doi.org/10.1103/PhysRevLett.116.186603} {\bibfield  {journal}
  {\bibinfo  {journal} {Phys.\ Rev.\ Lett.}\ }\textbf {\bibinfo {volume}
  {116}},\ \bibinfo {pages} {186603} (\bibinfo {year} {2016})}\BibitemShut
  {NoStop}%
\bibitem [{\citenamefont {Koch}\ and\ \citenamefont {von
  Oppen}(2005)}]{Koch2005}%
  \BibitemOpen
  \bibfield  {author} {\bibinfo {author} {\bibfnamefont {J.}~\bibnamefont
  {Koch}}\ and\ \bibinfo {author} {\bibfnamefont {F.}~\bibnamefont {von
  Oppen}},\ }\bibfield  {title} {\bibinfo {title} {Franck-condon blockade and
  giant fano factors in transport through single molecules},\ }\href
  {https://doi.org/10.1103/PhysRevLett.94.206804} {\bibfield  {journal}
  {\bibinfo  {journal} {Phys.\ Rev.\ Lett.}\ }\textbf {\bibinfo {volume}
  {94}},\ \bibinfo {pages} {206804} (\bibinfo {year} {2005})}\BibitemShut
  {NoStop}%
\bibitem [{\citenamefont {Leturcq}\ \emph {et~al.}(2009)\citenamefont
  {Leturcq}, \citenamefont {Stampfer}, \citenamefont {Inderbitzin},
  \citenamefont {Durrer}, \citenamefont {Hierold}, \citenamefont {Mariani},
  \citenamefont {Schultz}, \citenamefont {von Oppen},\ and\ \citenamefont
  {Ensslin}}]{Leturcq2009}%
  \BibitemOpen
  \bibfield  {author} {\bibinfo {author} {\bibfnamefont {R.}~\bibnamefont
  {Leturcq}}, \bibinfo {author} {\bibfnamefont {C.}~\bibnamefont {Stampfer}},
  \bibinfo {author} {\bibfnamefont {K.}~\bibnamefont {Inderbitzin}}, \bibinfo
  {author} {\bibfnamefont {L.}~\bibnamefont {Durrer}}, \bibinfo {author}
  {\bibfnamefont {C.}~\bibnamefont {Hierold}}, \bibinfo {author} {\bibfnamefont
  {E.}~\bibnamefont {Mariani}}, \bibinfo {author} {\bibfnamefont {M.~G.}\
  \bibnamefont {Schultz}}, \bibinfo {author} {\bibfnamefont {F.}~\bibnamefont
  {von Oppen}},\ and\ \bibinfo {author} {\bibfnamefont {K.}~\bibnamefont
  {Ensslin}},\ }\bibfield  {title} {\bibinfo {title} {Franck--condon blockade
  in suspended carbon nanotube quantum dots},\ }\href
  {https://doi.org/10.1038/nphys1234} {\bibfield  {journal} {\bibinfo
  {journal} {Nat.\ Phys.}\ }\textbf {\bibinfo {volume} {5}},\ \bibinfo {pages}
  {327} (\bibinfo {year} {2009})}\BibitemShut {NoStop}%
\bibitem [{\citenamefont {Kolli}\ \emph {et~al.}(2012)\citenamefont {Kolli},
  \citenamefont {O’Reilly}, \citenamefont {Scholes},\ and\ \citenamefont
  {Olaya-Castro}}]{Kolli2012}%
  \BibitemOpen
  \bibfield  {author} {\bibinfo {author} {\bibfnamefont {A.}~\bibnamefont
  {Kolli}}, \bibinfo {author} {\bibfnamefont {E.~J.}\ \bibnamefont
  {O’Reilly}}, \bibinfo {author} {\bibfnamefont {G.~D.}\ \bibnamefont
  {Scholes}},\ and\ \bibinfo {author} {\bibfnamefont {A.}~\bibnamefont
  {Olaya-Castro}},\ }\bibfield  {title} {\bibinfo {title} {The fundamental role
  of quantized vibrations in coherent light harvesting by cryptophyte algae},\
  }\href {https://doi.org/10.1063/1.4764100} {\bibfield  {journal} {\bibinfo
  {journal} {J.\ Chem.\ Phys.}\ }\textbf {\bibinfo {volume} {137}},\ \bibinfo
  {pages} {174109} (\bibinfo {year} {2012})}\BibitemShut {NoStop}%
\bibitem [{\citenamefont {Tiwari}\ \emph {et~al.}(2013)\citenamefont {Tiwari},
  \citenamefont {Peters},\ and\ \citenamefont {Jonas}}]{Tiwari2013}%
  \BibitemOpen
  \bibfield  {author} {\bibinfo {author} {\bibfnamefont {V.}~\bibnamefont
  {Tiwari}}, \bibinfo {author} {\bibfnamefont {W.~K.}\ \bibnamefont {Peters}},\
  and\ \bibinfo {author} {\bibfnamefont {D.~M.}\ \bibnamefont {Jonas}},\
  }\bibfield  {title} {\bibinfo {title} {Electronic resonance with
  anticorrelated pigment vibrations drives photosynthetic energy transfer
  outside the adiabatic framework},\ }\href
  {https://doi.org/10.1073/pnas.1211157110} {\bibfield  {journal} {\bibinfo
  {journal} {Proc.\ Natl.\ Acad.\ Sci.\ U.S.A.}\ }\textbf {\bibinfo {volume}
  {110}},\ \bibinfo {pages} {1203} (\bibinfo {year} {2013})}\BibitemShut
  {NoStop}%
\bibitem [{\citenamefont {Dijkstra}\ \emph {et~al.}(2015)\citenamefont
  {Dijkstra}, \citenamefont {Wang}, \citenamefont {Cao},\ and\ \citenamefont
  {Fleming}}]{Dijkstra}%
  \BibitemOpen
  \bibfield  {author} {\bibinfo {author} {\bibfnamefont {A.~G.}\ \bibnamefont
  {Dijkstra}}, \bibinfo {author} {\bibfnamefont {C.}~\bibnamefont {Wang}},
  \bibinfo {author} {\bibfnamefont {J.}~\bibnamefont {Cao}},\ and\ \bibinfo
  {author} {\bibfnamefont {G.~R.}\ \bibnamefont {Fleming}},\ }\bibfield
  {title} {\bibinfo {title} {Coherent exciton dynamics in the presence of
  underdamped vibrations},\ }\href {https://doi.org/10.1021/jz502701u}
  {\bibfield  {journal} {\bibinfo  {journal} {J.\ Phys.\ Chem.\ Lett.}\
  }\textbf {\bibinfo {volume} {6}},\ \bibinfo {pages} {627} (\bibinfo {year}
  {2015})}\BibitemShut {NoStop}%
\bibitem [{\citenamefont {Emin}\ and\ \citenamefont {Hart}(1987)}]{Emin1987}%
  \BibitemOpen
  \bibfield  {author} {\bibinfo {author} {\bibfnamefont {D.}~\bibnamefont
  {Emin}}\ and\ \bibinfo {author} {\bibfnamefont {C.~F.}\ \bibnamefont
  {Hart}},\ }\bibfield  {title} {\bibinfo {title} {Phonon-assisted hopping of
  an electron on a wannier-stark ladder in a strong electric field},\ }\href
  {https://doi.org/10.1103/PhysRevB.36.2530} {\bibfield  {journal} {\bibinfo
  {journal} {Phys.\ Rev.\ B}\ }\textbf {\bibinfo {volume} {36}},\ \bibinfo
  {pages} {2530} (\bibinfo {year} {1987})}\BibitemShut {NoStop}%
\bibitem [{\citenamefont {Moix}\ \emph {et~al.}(2013)\citenamefont {Moix},
  \citenamefont {Khasin},\ and\ \citenamefont {Cao}}]{cao150}%
  \BibitemOpen
  \bibfield  {author} {\bibinfo {author} {\bibfnamefont {J.~M.}\ \bibnamefont
  {Moix}}, \bibinfo {author} {\bibfnamefont {M.}~\bibnamefont {Khasin}},\ and\
  \bibinfo {author} {\bibfnamefont {J.}~\bibnamefont {Cao}},\ }\bibfield
  {title} {\bibinfo {title} {Coherent quantum transport in disordered systems:
  I.~the influence of dephasing on the transport properties and absorption
  spectra on one-dimensional systems},\ }\href
  {https://doi.org/10.1088/1367-2630/15/8/085010} {\bibfield  {journal}
  {\bibinfo  {journal} {New J.\ Phys.}\ }\textbf {\bibinfo {volume} {15}},\
  \bibinfo {pages} {085010} (\bibinfo {year} {2013})}\BibitemShut {NoStop}%
\bibitem [{\citenamefont {Sachdev}\ \emph {et~al.}(2002)\citenamefont
  {Sachdev}, \citenamefont {Sengupta},\ and\ \citenamefont
  {Girvin}}]{Sachdev2002}%
  \BibitemOpen
  \bibfield  {author} {\bibinfo {author} {\bibfnamefont {S.}~\bibnamefont
  {Sachdev}}, \bibinfo {author} {\bibfnamefont {K.}~\bibnamefont {Sengupta}},\
  and\ \bibinfo {author} {\bibfnamefont {S.~M.}\ \bibnamefont {Girvin}},\
  }\bibfield  {title} {\bibinfo {title} {Mott insulators in strong electric
  fields},\ }\href {https://doi.org/10.1103/PhysRevB.66.075128} {\bibfield
  {journal} {\bibinfo  {journal} {Phys.\ Rev.\ B}\ }\textbf {\bibinfo {volume}
  {66}},\ \bibinfo {pages} {075128} (\bibinfo {year} {2002})}\BibitemShut
  {NoStop}%
\bibitem [{\citenamefont {Sias}\ \emph {et~al.}(2007)\citenamefont {Sias},
  \citenamefont {Zenesini}, \citenamefont {Lignier}, \citenamefont {Wimberger},
  \citenamefont {Ciampini}, \citenamefont {Morsch},\ and\ \citenamefont
  {Arimondo}}]{Sias2007}%
  \BibitemOpen
  \bibfield  {author} {\bibinfo {author} {\bibfnamefont {C.}~\bibnamefont
  {Sias}}, \bibinfo {author} {\bibfnamefont {A.}~\bibnamefont {Zenesini}},
  \bibinfo {author} {\bibfnamefont {H.}~\bibnamefont {Lignier}}, \bibinfo
  {author} {\bibfnamefont {S.}~\bibnamefont {Wimberger}}, \bibinfo {author}
  {\bibfnamefont {D.}~\bibnamefont {Ciampini}}, \bibinfo {author}
  {\bibfnamefont {O.}~\bibnamefont {Morsch}},\ and\ \bibinfo {author}
  {\bibfnamefont {E.}~\bibnamefont {Arimondo}},\ }\bibfield  {title} {\bibinfo
  {title} {Resonantly enhanced tunneling of bose-einstein condensates in
  periodic potentials},\ }\href {https://doi.org/10.1103/PhysRevLett.98.120403}
  {\bibfield  {journal} {\bibinfo  {journal} {Phys.\ Rev.\ Lett.}\ }\textbf
  {\bibinfo {volume} {98}},\ \bibinfo {pages} {120403} (\bibinfo {year}
  {2007})}\BibitemShut {NoStop}%
\bibitem [{\citenamefont {Pielawa}\ \emph {et~al.}(2011)\citenamefont
  {Pielawa}, \citenamefont {Kitagawa}, \citenamefont {Berg},\ and\
  \citenamefont {Sachdev}}]{Pielawa2011}%
  \BibitemOpen
  \bibfield  {author} {\bibinfo {author} {\bibfnamefont {S.}~\bibnamefont
  {Pielawa}}, \bibinfo {author} {\bibfnamefont {T.}~\bibnamefont {Kitagawa}},
  \bibinfo {author} {\bibfnamefont {E.}~\bibnamefont {Berg}},\ and\ \bibinfo
  {author} {\bibfnamefont {S.}~\bibnamefont {Sachdev}},\ }\bibfield  {title}
  {\bibinfo {title} {Correlated phases of bosons in tilted frustrated
  lattices},\ }\href {https://doi.org/10.1103/PhysRevB.83.205135} {\bibfield
  {journal} {\bibinfo  {journal} {Phys. Rev. B}\ }\textbf {\bibinfo {volume}
  {83}},\ \bibinfo {pages} {205135} (\bibinfo {year} {2011})}\BibitemShut
  {NoStop}%
\bibitem [{\citenamefont {Simon}\ \emph {et~al.}(2011)\citenamefont {Simon},
  \citenamefont {Bakr}, \citenamefont {Ma}, \citenamefont {Tai}, \citenamefont
  {Preiss},\ and\ \citenamefont {Greiner}}]{Simon2011}%
  \BibitemOpen
  \bibfield  {author} {\bibinfo {author} {\bibfnamefont {J.}~\bibnamefont
  {Simon}}, \bibinfo {author} {\bibfnamefont {W.~S.}\ \bibnamefont {Bakr}},
  \bibinfo {author} {\bibfnamefont {R.}~\bibnamefont {Ma}}, \bibinfo {author}
  {\bibfnamefont {M.~E.}\ \bibnamefont {Tai}}, \bibinfo {author} {\bibfnamefont
  {P.~M.}\ \bibnamefont {Preiss}},\ and\ \bibinfo {author} {\bibfnamefont
  {M.}~\bibnamefont {Greiner}},\ }\bibfield  {title} {\bibinfo {title} {Quantum
  simulation of antiferromagnetic spin chains in an optical lattice},\ }\href
  {https://doi.org/10.1038/nature09994} {\bibfield  {journal} {\bibinfo
  {journal} {Nature}\ }\textbf {\bibinfo {volume} {472}},\ \bibinfo {pages}
  {307} (\bibinfo {year} {2011})}\BibitemShut {NoStop}%
\bibitem [{\citenamefont {Rubbo}\ \emph {et~al.}(2011)\citenamefont {Rubbo},
  \citenamefont {Manmana}, \citenamefont {Peden}, \citenamefont {Holland},\
  and\ \citenamefont {Rey}}]{Rubbo2011}%
  \BibitemOpen
  \bibfield  {author} {\bibinfo {author} {\bibfnamefont {C.~P.}\ \bibnamefont
  {Rubbo}}, \bibinfo {author} {\bibfnamefont {S.~R.}\ \bibnamefont {Manmana}},
  \bibinfo {author} {\bibfnamefont {B.~M.}\ \bibnamefont {Peden}}, \bibinfo
  {author} {\bibfnamefont {M.~J.}\ \bibnamefont {Holland}},\ and\ \bibinfo
  {author} {\bibfnamefont {A.~M.}\ \bibnamefont {Rey}},\ }\bibfield  {title}
  {\bibinfo {title} {Resonantly enhanced tunneling and transport of ultracold
  atoms on tilted optical lattices},\ }\href
  {https://doi.org/10.1103/PhysRevA.84.033638} {\bibfield  {journal} {\bibinfo
  {journal} {Phys.\ Rev.\ A}\ }\textbf {\bibinfo {volume} {84}},\ \bibinfo
  {pages} {033638} (\bibinfo {year} {2011})}\BibitemShut {NoStop}%
\bibitem [{\citenamefont {Carrasquilla}\ \emph {et~al.}(2013)\citenamefont
  {Carrasquilla}, \citenamefont {Manmana},\ and\ \citenamefont
  {Rigol}}]{Carrasquilla2013}%
  \BibitemOpen
  \bibfield  {author} {\bibinfo {author} {\bibfnamefont {J.}~\bibnamefont
  {Carrasquilla}}, \bibinfo {author} {\bibfnamefont {S.~R.}\ \bibnamefont
  {Manmana}},\ and\ \bibinfo {author} {\bibfnamefont {M.}~\bibnamefont
  {Rigol}},\ }\bibfield  {title} {\bibinfo {title} {Scaling of the gap,
  fidelity susceptibility, and bloch oscillations across the
  superfluid-to-mott-insulator transition in the one-dimensional bose-hubbard
  model},\ }\href {https://doi.org/10.1103/PhysRevA.87.043606} {\bibfield
  {journal} {\bibinfo  {journal} {Phys.\ Rev.\ A}\ }\textbf {\bibinfo {volume}
  {87}},\ \bibinfo {pages} {043606} (\bibinfo {year} {2013})}\BibitemShut
  {NoStop}%
\bibitem [{\citenamefont {Meinert}\ \emph {et~al.}(2013)\citenamefont
  {Meinert}, \citenamefont {Mark}, \citenamefont {Kirilov}, \citenamefont
  {Lauber}, \citenamefont {Weinmann}, \citenamefont {Daley},\ and\
  \citenamefont {N\"agerl}}]{Meinert2013}%
  \BibitemOpen
  \bibfield  {author} {\bibinfo {author} {\bibfnamefont {F.}~\bibnamefont
  {Meinert}}, \bibinfo {author} {\bibfnamefont {M.~J.}\ \bibnamefont {Mark}},
  \bibinfo {author} {\bibfnamefont {E.}~\bibnamefont {Kirilov}}, \bibinfo
  {author} {\bibfnamefont {K.}~\bibnamefont {Lauber}}, \bibinfo {author}
  {\bibfnamefont {P.}~\bibnamefont {Weinmann}}, \bibinfo {author}
  {\bibfnamefont {A.~J.}\ \bibnamefont {Daley}},\ and\ \bibinfo {author}
  {\bibfnamefont {H.-C.}\ \bibnamefont {N\"agerl}},\ }\bibfield  {title}
  {\bibinfo {title} {Quantum quench in an atomic one-dimensional ising chain},\
  }\href {https://doi.org/10.1103/PhysRevLett.111.053003} {\bibfield  {journal}
  {\bibinfo  {journal} {Phys.\ Rev.\ Lett.}\ }\textbf {\bibinfo {volume}
  {111}},\ \bibinfo {pages} {053003} (\bibinfo {year} {2013})}\BibitemShut
  {NoStop}%
\bibitem [{\citenamefont {Meinert}\ \emph {et~al.}(2014)\citenamefont
  {Meinert}, \citenamefont {Mark}, \citenamefont {Kirilov}, \citenamefont
  {Lauber}, \citenamefont {Weinmann}, \citenamefont {Gr{\"o}bner},
  \citenamefont {Daley},\ and\ \citenamefont {N{\"a}gerl}}]{Meinert2014}%
  \BibitemOpen
  \bibfield  {author} {\bibinfo {author} {\bibfnamefont {F.}~\bibnamefont
  {Meinert}}, \bibinfo {author} {\bibfnamefont {M.~J.}\ \bibnamefont {Mark}},
  \bibinfo {author} {\bibfnamefont {E.}~\bibnamefont {Kirilov}}, \bibinfo
  {author} {\bibfnamefont {K.}~\bibnamefont {Lauber}}, \bibinfo {author}
  {\bibfnamefont {P.}~\bibnamefont {Weinmann}}, \bibinfo {author}
  {\bibfnamefont {M.}~\bibnamefont {Gr{\"o}bner}}, \bibinfo {author}
  {\bibfnamefont {A.~J.}\ \bibnamefont {Daley}},\ and\ \bibinfo {author}
  {\bibfnamefont {H.-C.}\ \bibnamefont {N{\"a}gerl}},\ }\bibfield  {title}
  {\bibinfo {title} {Observation of many-body dynamics in long-range tunneling
  after a quantum quench},\ }\href {https://doi.org/10.1126/science.1248402}
  {\bibfield  {journal} {\bibinfo  {journal} {Science}\ }\textbf {\bibinfo
  {volume} {344}},\ \bibinfo {pages} {1259} (\bibinfo {year}
  {2014})}\BibitemShut {NoStop}%
\bibitem [{\citenamefont {Buyskikh}\ \emph
  {et~al.}(2019{\natexlab{a}})\citenamefont {Buyskikh}, \citenamefont
  {Tagliacozzo}, \citenamefont {Schuricht}, \citenamefont {Hooley},
  \citenamefont {Pekker},\ and\ \citenamefont {Daley}}]{Buyskikh2019}%
  \BibitemOpen
  \bibfield  {author} {\bibinfo {author} {\bibfnamefont {A.~S.}\ \bibnamefont
  {Buyskikh}}, \bibinfo {author} {\bibfnamefont {L.}~\bibnamefont
  {Tagliacozzo}}, \bibinfo {author} {\bibfnamefont {D.}~\bibnamefont
  {Schuricht}}, \bibinfo {author} {\bibfnamefont {C.~A.}\ \bibnamefont
  {Hooley}}, \bibinfo {author} {\bibfnamefont {D.}~\bibnamefont {Pekker}},\
  and\ \bibinfo {author} {\bibfnamefont {A.~J.}\ \bibnamefont {Daley}},\
  }\bibfield  {title} {\bibinfo {title} {Spin models, dynamics, and criticality
  with atoms in tilted optical superlattices},\ }\href
  {https://doi.org/10.1103/PhysRevLett.123.090401} {\bibfield  {journal}
  {\bibinfo  {journal} {Phys.\ Rev.\ Lett.}\ }\textbf {\bibinfo {volume}
  {123}},\ \bibinfo {pages} {090401} (\bibinfo {year}
  {2019}{\natexlab{a}})}\BibitemShut {NoStop}%
\bibitem [{\citenamefont {Buyskikh}\ \emph
  {et~al.}(2019{\natexlab{b}})\citenamefont {Buyskikh}, \citenamefont
  {Tagliacozzo}, \citenamefont {Schuricht}, \citenamefont {Hooley},
  \citenamefont {Pekker},\ and\ \citenamefont {Daley}}]{Buyskikh2019_2}%
  \BibitemOpen
  \bibfield  {author} {\bibinfo {author} {\bibfnamefont {A.~S.}\ \bibnamefont
  {Buyskikh}}, \bibinfo {author} {\bibfnamefont {L.}~\bibnamefont
  {Tagliacozzo}}, \bibinfo {author} {\bibfnamefont {D.}~\bibnamefont
  {Schuricht}}, \bibinfo {author} {\bibfnamefont {C.~A.}\ \bibnamefont
  {Hooley}}, \bibinfo {author} {\bibfnamefont {D.}~\bibnamefont {Pekker}},\
  and\ \bibinfo {author} {\bibfnamefont {A.~J.}\ \bibnamefont {Daley}},\
  }\bibfield  {title} {\bibinfo {title} {Resonant two-site tunneling dynamics
  of bosons in a tilted optical superlattice},\ }\href
  {https://doi.org/10.1103/PhysRevA.100.023627} {\bibfield  {journal} {\bibinfo
   {journal} {Phys.\ Rev.\ A}\ }\textbf {\bibinfo {volume} {100}},\ \bibinfo
  {pages} {023627} (\bibinfo {year} {2019}{\natexlab{b}})}\BibitemShut
  {NoStop}%
\bibitem [{\citenamefont {Gorshkov}\ \emph {et~al.}(2011)\citenamefont
  {Gorshkov}, \citenamefont {Manmana}, \citenamefont {Chen}, \citenamefont
  {Ye}, \citenamefont {Demler}, \citenamefont {Lukin},\ and\ \citenamefont
  {Rey}}]{Gorshkov2011}%
  \BibitemOpen
  \bibfield  {author} {\bibinfo {author} {\bibfnamefont {A.~V.}\ \bibnamefont
  {Gorshkov}}, \bibinfo {author} {\bibfnamefont {S.~R.}\ \bibnamefont
  {Manmana}}, \bibinfo {author} {\bibfnamefont {G.}~\bibnamefont {Chen}},
  \bibinfo {author} {\bibfnamefont {J.}~\bibnamefont {Ye}}, \bibinfo {author}
  {\bibfnamefont {E.}~\bibnamefont {Demler}}, \bibinfo {author} {\bibfnamefont
  {M.~D.}\ \bibnamefont {Lukin}},\ and\ \bibinfo {author} {\bibfnamefont
  {A.~M.}\ \bibnamefont {Rey}},\ }\bibfield  {title} {\bibinfo {title} {Tunable
  superfluidity and quantum magnetism with ultracold polar molecules},\ }\href
  {https://doi.org/10.1103/PhysRevLett.107.115301} {\bibfield  {journal}
  {\bibinfo  {journal} {Phys.\ Rev.\ Lett.}\ }\textbf {\bibinfo {volume}
  {107}},\ \bibinfo {pages} {115301} (\bibinfo {year} {2011})}\BibitemShut
  {NoStop}%
\bibitem [{\citenamefont {Keilmann}\ \emph {et~al.}(2011)\citenamefont
  {Keilmann}, \citenamefont {Lanzmich}, \citenamefont {McCulloch},\ and\
  \citenamefont {Roncaglia}}]{Keilmann2011}%
  \BibitemOpen
  \bibfield  {author} {\bibinfo {author} {\bibfnamefont {T.}~\bibnamefont
  {Keilmann}}, \bibinfo {author} {\bibfnamefont {S.}~\bibnamefont {Lanzmich}},
  \bibinfo {author} {\bibfnamefont {I.}~\bibnamefont {McCulloch}},\ and\
  \bibinfo {author} {\bibfnamefont {M.}~\bibnamefont {Roncaglia}},\ }\bibfield
  {title} {\bibinfo {title} {Statistically induced phase transitions and anyons
  in 1d optical lattices},\ }\href {https://doi.org/10.1038/ncomms1353}
  {\bibfield  {journal} {\bibinfo  {journal} {Nat.\ Commun.}\ }\textbf
  {\bibinfo {volume} {2}},\ \bibinfo {pages} {361} (\bibinfo {year}
  {2011})}\BibitemShut {NoStop}%
\bibitem [{\citenamefont {Kloss}\ \emph {et~al.}(2019)\citenamefont {Kloss},
  \citenamefont {Reichman},\ and\ \citenamefont {Tempelaar}}]{Kloss}%
  \BibitemOpen
  \bibfield  {author} {\bibinfo {author} {\bibfnamefont {B.}~\bibnamefont
  {Kloss}}, \bibinfo {author} {\bibfnamefont {D.~R.}\ \bibnamefont
  {Reichman}},\ and\ \bibinfo {author} {\bibfnamefont {R.}~\bibnamefont
  {Tempelaar}},\ }\bibfield  {title} {\bibinfo {title} {Multiset matrix product
  state calculations reveal mobile franck--condon excitations under strong
  holstein-type coupling},\ }\href
  {https://doi.org/10.1103/PhysRevLett.123.126601} {\bibfield  {journal}
  {\bibinfo  {journal} {Phys.\ Rev.\ Lett.}\ }\textbf {\bibinfo {volume}
  {123}},\ \bibinfo {pages} {126601} (\bibinfo {year} {2019})}\BibitemShut
  {NoStop}%
\bibitem [{\citenamefont {Shalashilin}\ and\ \citenamefont
  {Child}(2000)}]{Shalashilin2000}%
  \BibitemOpen
  \bibfield  {author} {\bibinfo {author} {\bibfnamefont {D.~V.}\ \bibnamefont
  {Shalashilin}}\ and\ \bibinfo {author} {\bibfnamefont {M.~S.}\ \bibnamefont
  {Child}},\ }\bibfield  {title} {\bibinfo {title} {Time dependent quantum
  propagation in phase space},\ }\href {https://doi.org/10.1063/1.1322075}
  {\bibfield  {journal} {\bibinfo  {journal} {J.\ Chem.\ Phys.}\ }\textbf
  {\bibinfo {volume} {113}},\ \bibinfo {pages} {10028} (\bibinfo {year}
  {2000})}\BibitemShut {NoStop}%
\bibitem [{\citenamefont {Shalashilin}\ and\ \citenamefont
  {Child}(2004)}]{Shalashilin2006}%
  \BibitemOpen
  \bibfield  {author} {\bibinfo {author} {\bibfnamefont {D.~V.}\ \bibnamefont
  {Shalashilin}}\ and\ \bibinfo {author} {\bibfnamefont {M.~S.}\ \bibnamefont
  {Child}},\ }\bibfield  {title} {\bibinfo {title} {Real time quantum
  propagation on a monte carlo trajectory guided grids of coupled coherent
  states: 26d simulation of pyrazine absorption spectrum},\ }\href
  {https://doi.org/10.1063/1.1776111} {\bibfield  {journal} {\bibinfo
  {journal} {J.\ Chem.\ Phys.}\ }\textbf {\bibinfo {volume} {121}},\ \bibinfo
  {pages} {3563} (\bibinfo {year} {2004})}\BibitemShut {NoStop}%
\bibitem [{\citenamefont {Ben-Nun}\ \emph {et~al.}(2000)\citenamefont
  {Ben-Nun}, \citenamefont {Quenneville},\ and\ \citenamefont
  {Martínez}}]{Ben-Nun}%
  \BibitemOpen
  \bibfield  {author} {\bibinfo {author} {\bibfnamefont {M.}~\bibnamefont
  {Ben-Nun}}, \bibinfo {author} {\bibfnamefont {J.}~\bibnamefont
  {Quenneville}},\ and\ \bibinfo {author} {\bibfnamefont {T.~J.}\ \bibnamefont
  {Martínez}},\ }\bibfield  {title} {\bibinfo {title} {Ab initio multiple
  spawning: Photochemistry from first principles quantum molecular dynamics},\
  }\href {https://doi.org/10.1021/jp994174i} {\bibfield  {journal} {\bibinfo
  {journal} {J.\ Phys.\ Chem.}\ }\textbf {\bibinfo {volume} {104}},\ \bibinfo
  {pages} {5161} (\bibinfo {year} {2000})}\BibitemShut {NoStop}%
\bibitem [{\citenamefont {Gu}\ and\ \citenamefont {Garashchuk}(2016)}]{Gu2016}%
  \BibitemOpen
  \bibfield  {author} {\bibinfo {author} {\bibfnamefont {B.}~\bibnamefont
  {Gu}}\ and\ \bibinfo {author} {\bibfnamefont {S.}~\bibnamefont
  {Garashchuk}},\ }\bibfield  {title} {\bibinfo {title} {Quantum dynamics with
  gaussian bases defined by the quantum trajectories},\ }\href
  {https://doi.org/10.1021/acs.jpca.5b10029} {\bibfield  {journal} {\bibinfo
  {journal} {The Journal of Physical Chemistry A}\ }\textbf {\bibinfo {volume}
  {120}},\ \bibinfo {pages} {3023} (\bibinfo {year} {2016})}\BibitemShut
  {NoStop}%
\bibitem [{\citenamefont {Werther}\ and\ \citenamefont
  {Gro\ss{}mann}(2020)}]{Werther2020}%
  \BibitemOpen
  \bibfield  {author} {\bibinfo {author} {\bibfnamefont {M.}~\bibnamefont
  {Werther}}\ and\ \bibinfo {author} {\bibfnamefont {F.}~\bibnamefont
  {Gro\ss{}mann}},\ }\bibfield  {title} {\bibinfo {title} {Apoptosis of moving
  nonorthogonal basis functions in many-particle quantum dynamics},\ }\href
  {https://doi.org/10.1103/PhysRevB.101.174315} {\bibfield  {journal} {\bibinfo
   {journal} {Phys.\ Rev.\ B}\ }\textbf {\bibinfo {volume} {101}},\ \bibinfo
  {pages} {174315} (\bibinfo {year} {2020})}\BibitemShut {NoStop}%
\bibitem [{\citenamefont {Koch}\ and\ \citenamefont
  {Frankcombe}(2013)}]{Koch2013}%
  \BibitemOpen
  \bibfield  {author} {\bibinfo {author} {\bibfnamefont {W.}~\bibnamefont
  {Koch}}\ and\ \bibinfo {author} {\bibfnamefont {T.~J.}\ \bibnamefont
  {Frankcombe}},\ }\bibfield  {title} {\bibinfo {title} {Basis expansion
  leaping: A new method to solve the time-dependent schr\"odinger equation for
  molecular quantum dynamics},\ }\href
  {https://doi.org/10.1103/PhysRevLett.110.263202} {\bibfield  {journal}
  {\bibinfo  {journal} {Phys.\ Rev.\ Lett.}\ }\textbf {\bibinfo {volume}
  {110}},\ \bibinfo {pages} {263202} (\bibinfo {year} {2013})}\BibitemShut
  {NoStop}%
\bibitem [{\citenamefont {Richings}\ \emph {et~al.}(2015)\citenamefont
  {Richings}, \citenamefont {Polyak}, \citenamefont {Spinlove}, \citenamefont
  {Worth}, \citenamefont {Burghardt},\ and\ \citenamefont
  {Lasorne}}]{Richings}%
  \BibitemOpen
  \bibfield  {author} {\bibinfo {author} {\bibfnamefont {G.}~\bibnamefont
  {Richings}}, \bibinfo {author} {\bibfnamefont {I.}~\bibnamefont {Polyak}},
  \bibinfo {author} {\bibfnamefont {K.}~\bibnamefont {Spinlove}}, \bibinfo
  {author} {\bibfnamefont {G.}~\bibnamefont {Worth}}, \bibinfo {author}
  {\bibfnamefont {I.}~\bibnamefont {Burghardt}},\ and\ \bibinfo {author}
  {\bibfnamefont {B.}~\bibnamefont {Lasorne}},\ }\bibfield  {title} {\bibinfo
  {title} {Quantum dynamics simulations using gaussian wavepackets: the vmcg
  method},\ }\href {https://doi.org/10.1080/0144235X.2015.1051354} {\bibfield
  {journal} {\bibinfo  {journal} {Int.\ Rev.\ Phys.\ Chem.}\ }\textbf {\bibinfo
  {volume} {34}},\ \bibinfo {pages} {269} (\bibinfo {year} {2015})}\BibitemShut
  {NoStop}%
\bibitem [{\citenamefont {Saller}\ and\ \citenamefont
  {Habershon}(2017)}]{Saller2017}%
  \BibitemOpen
  \bibfield  {author} {\bibinfo {author} {\bibfnamefont {M.~A.~C.}\
  \bibnamefont {Saller}}\ and\ \bibinfo {author} {\bibfnamefont
  {S.}~\bibnamefont {Habershon}},\ }\bibfield  {title} {\bibinfo {title}
  {Quantum dynamics with short-time trajectories and minimal adaptive basis
  sets},\ }\href {https://doi.org/10.1021/acs.jctc.7b00021} {\bibfield
  {journal} {\bibinfo  {journal} {J.\ Chem.\ Theory Comput.}\ }\textbf
  {\bibinfo {volume} {13}},\ \bibinfo {pages} {3085} (\bibinfo {year}
  {2017})}\BibitemShut {NoStop}%
\bibitem [{\citenamefont {Hartke}(2006)}]{Hartke2006}%
  \BibitemOpen
  \bibfield  {author} {\bibinfo {author} {\bibfnamefont {B.}~\bibnamefont
  {Hartke}},\ }\bibfield  {title} {\bibinfo {title} {Propagation with
  distributed gaussians as a sparse{,} adaptive basis for higher-dimensional
  quantum dynamics},\ }\href {https://doi.org/10.1039/B606376D} {\bibfield
  {journal} {\bibinfo  {journal} {Phys.\ Chem.\ Chem.\ Phys.}\ }\textbf
  {\bibinfo {volume} {8}},\ \bibinfo {pages} {3627} (\bibinfo {year}
  {2006})}\BibitemShut {NoStop}%
\bibitem [{\citenamefont {Sielk}\ \emph {et~al.}(2009)\citenamefont {Sielk},
  \citenamefont {von Horsten}, \citenamefont {Krüger}, \citenamefont
  {Schneider},\ and\ \citenamefont {Hartke}}]{Sielk2009}%
  \BibitemOpen
  \bibfield  {author} {\bibinfo {author} {\bibfnamefont {J.}~\bibnamefont
  {Sielk}}, \bibinfo {author} {\bibfnamefont {H.~F.}\ \bibnamefont {von
  Horsten}}, \bibinfo {author} {\bibfnamefont {F.}~\bibnamefont {Krüger}},
  \bibinfo {author} {\bibfnamefont {R.}~\bibnamefont {Schneider}},\ and\
  \bibinfo {author} {\bibfnamefont {B.}~\bibnamefont {Hartke}},\ }\bibfield
  {title} {\bibinfo {title} {Quantum-mechanical wavepacket propagation in a
  sparse{,} adaptive basis of interpolating gaussians with collocation},\
  }\href {https://doi.org/10.1039/B814315C} {\bibfield  {journal} {\bibinfo
  {journal} {Phys.\ Chem.\ Chem.\ Phys.}\ }\textbf {\bibinfo {volume} {11}},\
  \bibinfo {pages} {463} (\bibinfo {year} {2009})}\BibitemShut {NoStop}%
\bibitem [{\citenamefont {Bon\v{c}a}\ \emph {et~al.}(1999)\citenamefont
  {Bon\v{c}a}, \citenamefont {Trugman},\ and\ \citenamefont
  {Batisti\'{c}}}]{Bonca}%
  \BibitemOpen
  \bibfield  {author} {\bibinfo {author} {\bibfnamefont {J.}~\bibnamefont
  {Bon\v{c}a}}, \bibinfo {author} {\bibfnamefont {S.~A.}\ \bibnamefont
  {Trugman}},\ and\ \bibinfo {author} {\bibfnamefont {I.}~\bibnamefont
  {Batisti\'{c}}},\ }\bibfield  {title} {\bibinfo {title} {Holstein polaron},\
  }\href {https://doi.org/10.1103/PhysRevB.60.1633} {\bibfield  {journal}
  {\bibinfo  {journal} {Phys.\ Rev.\ B}\ }\textbf {\bibinfo {volume} {60}},\
  \bibinfo {pages} {1633} (\bibinfo {year} {1999})}\BibitemShut {NoStop}%
\bibitem [{\citenamefont {Dorfner}\ \emph {et~al.}(2015)\citenamefont
  {Dorfner}, \citenamefont {Vidmar}, \citenamefont {Brockt}, \citenamefont
  {Jeckelmann},\ and\ \citenamefont {Heidrich-Meisner}}]{FHM2015}%
  \BibitemOpen
  \bibfield  {author} {\bibinfo {author} {\bibfnamefont {F.}~\bibnamefont
  {Dorfner}}, \bibinfo {author} {\bibfnamefont {L.}~\bibnamefont {Vidmar}},
  \bibinfo {author} {\bibfnamefont {C.}~\bibnamefont {Brockt}}, \bibinfo
  {author} {\bibfnamefont {E.}~\bibnamefont {Jeckelmann}},\ and\ \bibinfo
  {author} {\bibfnamefont {F.}~\bibnamefont {Heidrich-Meisner}},\ }\bibfield
  {title} {\bibinfo {title} {Real-time decay of a highly excited charge carrier
  in the one-dimensional holstein model},\ }\href
  {https://doi.org/10.1103/PhysRevB.91.104302} {\bibfield  {journal} {\bibinfo
  {journal} {Phys.\ Rev.\ B}\ }\textbf {\bibinfo {volume} {91}},\ \bibinfo
  {pages} {104302} (\bibinfo {year} {2015})}\BibitemShut {NoStop}%
\bibitem [{Note1()}]{Note1}%
  \BibitemOpen
  \bibinfo {note} {Based on our mechanistic explanation, we can expect to find
  the same resonance behavior and multimodal peaked states for any arbitrarily
  large integers $m$ and $n$ with $\protect \frac {m}{n} = \delta _\epsilon
  /(\hbar \omega )$. However, the necessary hopping strength $\abs {J}$ and the
  required system size to observe the resonance effects increase with $n$;
  thus, for finite system sizes and finite $J$, only small values of $n$ lead
  to observable resonance behavior. Furthermore, the larger $m$, the smaller
  the number of resonant $\ket {\nu '}$ states in the uphill direction, leading
  to more pronounced asymmetry in the tunneling.}\BibitemShut {Stop}%
\bibitem [{\citenamefont {Glauber}(1963)}]{Glauber}%
  \BibitemOpen
  \bibfield  {author} {\bibinfo {author} {\bibfnamefont {R.~J.}\ \bibnamefont
  {Glauber}},\ }\bibfield  {title} {\bibinfo {title} {Coherent and incoherent
  states of the radiation field},\ }\href
  {https://doi.org/10.1103/PhysRev.131.2766} {\bibfield  {journal} {\bibinfo
  {journal} {Phys.\ Rev.}\ }\textbf {\bibinfo {volume} {131}},\ \bibinfo
  {pages} {2766} (\bibinfo {year} {1963})}\BibitemShut {NoStop}%
\bibitem [{\citenamefont {Cao}\ and\ \citenamefont {Silbey}(2009)}]{cao104}%
  \BibitemOpen
  \bibfield  {author} {\bibinfo {author} {\bibfnamefont {J.}~\bibnamefont
  {Cao}}\ and\ \bibinfo {author} {\bibfnamefont {R.~J.}\ \bibnamefont
  {Silbey}},\ }\bibfield  {title} {\bibinfo {title} {Optimization of exciton
  trapping in energy transfer processes},\ }\href
  {https://doi.org/10.1021/jp9032589} {\bibfield  {journal} {\bibinfo
  {journal} {J.\ Phys.\ Chem.\ A}\ }\textbf {\bibinfo {volume} {113}},\
  \bibinfo {pages} {13825} (\bibinfo {year} {2009})}\BibitemShut {NoStop}%
\bibitem [{\citenamefont {Wu}\ and\ \citenamefont {Cao}(2013)}]{cao123}%
  \BibitemOpen
  \bibfield  {author} {\bibinfo {author} {\bibfnamefont {J.}~\bibnamefont
  {Wu}}\ and\ \bibinfo {author} {\bibfnamefont {J.}~\bibnamefont {Cao}},\
  }\bibfield  {title} {\bibinfo {title} {Higher-order kinetic expansion of
  quantum dissipative dynamics: Mapping quantum networks to kinetic networks},\
  }\href {https://doi.org/10.1063/1.4812781} {\bibfield  {journal} {\bibinfo
  {journal} {J.\ Chem.\ Phys.}\ }\textbf {\bibinfo {volume} {139}},\ \bibinfo
  {pages} {044102} (\bibinfo {year} {2013})}\BibitemShut {NoStop}%
\bibitem [{\citenamefont {Cerrillo}\ and\ \citenamefont
  {Cao}(2014)}]{Cerrillo2014}%
  \BibitemOpen
  \bibfield  {author} {\bibinfo {author} {\bibfnamefont {J.}~\bibnamefont
  {Cerrillo}}\ and\ \bibinfo {author} {\bibfnamefont {J.}~\bibnamefont {Cao}},\
  }\bibfield  {title} {\bibinfo {title} {Non-markovian dynamical maps:
  Numerical processing of open quantum trajectories},\ }\href
  {https://doi.org/10.1103/PhysRevLett.112.110401} {\bibfield  {journal}
  {\bibinfo  {journal} {Phys. Rev. Lett.}\ }\textbf {\bibinfo {volume} {112}},\
  \bibinfo {pages} {110401} (\bibinfo {year} {2014})}\BibitemShut {NoStop}%
\bibitem [{\citenamefont {Shlesinger}(1974)}]{Shlesinger1974}%
  \BibitemOpen
  \bibfield  {author} {\bibinfo {author} {\bibfnamefont {M.~F.}\ \bibnamefont
  {Shlesinger}},\ }\bibfield  {title} {\bibinfo {title} {Asymptotic solutions
  of continuous-time random walks},\ }\href
  {https://doi.org/10.1007/BF01008803} {\bibfield  {journal} {\bibinfo
  {journal} {J.\ Stat.\ Phys.}\ }\textbf {\bibinfo {volume} {10}},\ \bibinfo
  {pages} {421} (\bibinfo {year} {1974})}\BibitemShut {NoStop}%
\bibitem [{\citenamefont {Witkoskie}\ and\ \citenamefont {Cao}(2006)}]{cao089}%
  \BibitemOpen
  \bibfield  {author} {\bibinfo {author} {\bibfnamefont {J.~B.}\ \bibnamefont
  {Witkoskie}}\ and\ \bibinfo {author} {\bibfnamefont {J.}~\bibnamefont
  {Cao}},\ }\bibfield  {title} {\bibinfo {title} {Aging correlation functions
  of the interrupted fractional fokker-planck propagator},\ }\href
  {https://doi.org/10.1063/1.2403874} {\bibfield  {journal} {\bibinfo
  {journal} {J.\ Chem.\ Phys.}\ }\textbf {\bibinfo {volume} {125}},\ \bibinfo
  {pages} {244511} (\bibinfo {year} {2006})}\BibitemShut {NoStop}%
\bibitem [{\citenamefont {Hart}\ \emph {et~al.}(2021)\citenamefont {Hart},
  \citenamefont {Chen}, \citenamefont {Banal}, \citenamefont {Bricker},
  \citenamefont {Dodin}, \citenamefont {Markova}, \citenamefont {Vyborna},
  \citenamefont {Willard}, \citenamefont {Häner}, \citenamefont {Bathe},\ and\
  \citenamefont {Schlau-Cohen}}]{Hart2021}%
  \BibitemOpen
  \bibfield  {author} {\bibinfo {author} {\bibfnamefont {S.~M.}\ \bibnamefont
  {Hart}}, \bibinfo {author} {\bibfnamefont {W.~J.}\ \bibnamefont {Chen}},
  \bibinfo {author} {\bibfnamefont {J.~L.}\ \bibnamefont {Banal}}, \bibinfo
  {author} {\bibfnamefont {W.~P.}\ \bibnamefont {Bricker}}, \bibinfo {author}
  {\bibfnamefont {A.}~\bibnamefont {Dodin}}, \bibinfo {author} {\bibfnamefont
  {L.}~\bibnamefont {Markova}}, \bibinfo {author} {\bibfnamefont
  {Y.}~\bibnamefont {Vyborna}}, \bibinfo {author} {\bibfnamefont {A.~P.}\
  \bibnamefont {Willard}}, \bibinfo {author} {\bibfnamefont {R.}~\bibnamefont
  {Häner}}, \bibinfo {author} {\bibfnamefont {M.}~\bibnamefont {Bathe}},\ and\
  \bibinfo {author} {\bibfnamefont {G.~S.}\ \bibnamefont {Schlau-Cohen}},\
  }\bibfield  {title} {\bibinfo {title} {Engineering couplings for exciton
  transport using synthetic dna scaffolds},\ }\href
  {https://doi.org/https://doi.org/10.1016/j.chempr.2020.12.020} {\bibfield
  {journal} {\bibinfo  {journal} {Chem}\ }\textbf {\bibinfo {volume} {7}},\
  \bibinfo {pages} {752} (\bibinfo {year} {2021})}\BibitemShut {NoStop}%
\bibitem [{\citenamefont {Hart}\ \emph {et~al.}(2022)\citenamefont {Hart},
  \citenamefont {Wang}, \citenamefont {Guo}, \citenamefont {Bathe},\ and\
  \citenamefont {Schlau-Cohen}}]{Hart2022}%
  \BibitemOpen
  \bibfield  {author} {\bibinfo {author} {\bibfnamefont {S.~M.}\ \bibnamefont
  {Hart}}, \bibinfo {author} {\bibfnamefont {X.}~\bibnamefont {Wang}}, \bibinfo
  {author} {\bibfnamefont {J.}~\bibnamefont {Guo}}, \bibinfo {author}
  {\bibfnamefont {M.}~\bibnamefont {Bathe}},\ and\ \bibinfo {author}
  {\bibfnamefont {G.~S.}\ \bibnamefont {Schlau-Cohen}},\ }\bibfield  {title}
  {\bibinfo {title} {Tuning optical absorption and emission using strongly
  coupled dimers in programmable dna scaffolds},\ }\href
  {https://doi.org/10.1021/acs.jpclett.1c03848} {\bibfield  {journal} {\bibinfo
   {journal} {J.\ Phys.\ Chem.\ Lett.}\ }\textbf {\bibinfo {volume} {13}},\
  \bibinfo {pages} {1863} (\bibinfo {year} {2022})}\BibitemShut {NoStop}%
\bibitem [{\citenamefont {van Nieuwenburg}\ \emph {et~al.}(2019)\citenamefont
  {van Nieuwenburg}, \citenamefont {Baum},\ and\ \citenamefont
  {Refael}}]{Nieuwenburg2019}%
  \BibitemOpen
  \bibfield  {author} {\bibinfo {author} {\bibfnamefont {E.}~\bibnamefont {van
  Nieuwenburg}}, \bibinfo {author} {\bibfnamefont {Y.}~\bibnamefont {Baum}},\
  and\ \bibinfo {author} {\bibfnamefont {G.}~\bibnamefont {Refael}},\
  }\bibfield  {title} {\bibinfo {title} {From bloch oscillations to many-body
  localization in clean interacting systems},\ }\href
  {https://doi.org/10.1073/pnas.1819316116} {\bibfield  {journal} {\bibinfo
  {journal} {Proc.\ Natl.\ Acad.\ Sci.\ U.S.A.}\ }\textbf {\bibinfo {volume}
  {116}},\ \bibinfo {pages} {9269} (\bibinfo {year} {2019})},\ \Eprint
  {https://arxiv.org/abs/https://www.pnas.org/doi/pdf/10.1073/pnas.1819316116}
  {https://www.pnas.org/doi/pdf/10.1073/pnas.1819316116} \BibitemShut {NoStop}%
\bibitem [{\citenamefont {Bleuse}\ \emph {et~al.}(1988)\citenamefont {Bleuse},
  \citenamefont {Bastard},\ and\ \citenamefont {Voisin}}]{Bleuse1988}%
  \BibitemOpen
  \bibfield  {author} {\bibinfo {author} {\bibfnamefont {J.}~\bibnamefont
  {Bleuse}}, \bibinfo {author} {\bibfnamefont {G.}~\bibnamefont {Bastard}},\
  and\ \bibinfo {author} {\bibfnamefont {P.}~\bibnamefont {Voisin}},\
  }\bibfield  {title} {\bibinfo {title} {Electric-field-induced localization
  and oscillatory electro-optical properties of semiconductor superlattices},\
  }\href {https://doi.org/10.1103/PhysRevLett.60.220} {\bibfield  {journal}
  {\bibinfo  {journal} {Phys.\ Rev.\ Lett.}\ }\textbf {\bibinfo {volume}
  {60}},\ \bibinfo {pages} {220} (\bibinfo {year} {1988})}\BibitemShut
  {NoStop}%
\bibitem [{\citenamefont {Mendez}\ \emph {et~al.}(1988)\citenamefont {Mendez},
  \citenamefont {Agull\'o-Rueda},\ and\ \citenamefont {Hong}}]{Mendez1988}%
  \BibitemOpen
  \bibfield  {author} {\bibinfo {author} {\bibfnamefont {E.~E.}\ \bibnamefont
  {Mendez}}, \bibinfo {author} {\bibfnamefont {F.}~\bibnamefont
  {Agull\'o-Rueda}},\ and\ \bibinfo {author} {\bibfnamefont {J.~M.}\
  \bibnamefont {Hong}},\ }\bibfield  {title} {\bibinfo {title} {Stark
  localization in gaas-gaalas superlattices under an electric field},\ }\href
  {https://doi.org/10.1103/PhysRevLett.60.2426} {\bibfield  {journal} {\bibinfo
   {journal} {Phys.\ Rev.\ Lett.}\ }\textbf {\bibinfo {volume} {60}},\ \bibinfo
  {pages} {2426} (\bibinfo {year} {1988})}\BibitemShut {NoStop}%
\bibitem [{\citenamefont {Weiser}\ \emph {et~al.}(1992)\citenamefont {Weiser},
  \citenamefont {Weihofen}, \citenamefont {Perales},\ and\ \citenamefont
  {Starck}}]{Weiser1992}%
  \BibitemOpen
  \bibfield  {author} {\bibinfo {author} {\bibfnamefont {G.}~\bibnamefont
  {Weiser}}, \bibinfo {author} {\bibfnamefont {R.}~\bibnamefont {Weihofen}},
  \bibinfo {author} {\bibfnamefont {A.}~\bibnamefont {Perales}},\ and\ \bibinfo
  {author} {\bibfnamefont {C.}~\bibnamefont {Starck}},\ }\bibfield  {title}
  {\bibinfo {title} {Stark effect and wannier-stark localization in ingaas
  quantum wells},\ }in\ \href {https://doi.org/10.1109/ICIPRM.1992.235621}
  {\emph {\bibinfo {booktitle} {LEOS 1992 Summer Topical Meeting Digest on
  Broadband Analog and Digital Optoelectronics, Optical Multiple Access
  Networks, Integrated Optoelectronics, and Smart Pixels}}}\ (\bibinfo {year}
  {1992})\ pp.\ \bibinfo {pages} {580--583}\BibitemShut {NoStop}%
\bibitem [{\citenamefont {Chin}\ \emph {et~al.}(2012)\citenamefont {Chin},
  \citenamefont {Huelga},\ and\ \citenamefont {Plenio}}]{Chin2012}%
  \BibitemOpen
  \bibfield  {author} {\bibinfo {author} {\bibfnamefont {A.~W.}\ \bibnamefont
  {Chin}}, \bibinfo {author} {\bibfnamefont {S.~F.}\ \bibnamefont {Huelga}},\
  and\ \bibinfo {author} {\bibfnamefont {M.~B.}\ \bibnamefont {Plenio}},\
  }\bibfield  {title} {\bibinfo {title} {Coherence and decoherence in
  biological systems: Principles of noise-assisted transport and the origin of
  long-lived coherences},\ }\href {https://doi.org/10.1098/rsta.2011.0224}
  {\bibfield  {journal} {\bibinfo  {journal} {Philos.\ Trans.\ R.\ Soc.\ A}\
  }\textbf {\bibinfo {volume} {370}},\ \bibinfo {pages} {3638} (\bibinfo {year}
  {2012})}\BibitemShut {NoStop}%
\bibitem [{\citenamefont {Rey}\ \emph {et~al.}(2013)\citenamefont {Rey},
  \citenamefont {Chin}, \citenamefont {Huelga},\ and\ \citenamefont
  {Plenio}}]{delRey2013}%
  \BibitemOpen
  \bibfield  {author} {\bibinfo {author} {\bibfnamefont {M.~d.}\ \bibnamefont
  {Rey}}, \bibinfo {author} {\bibfnamefont {A.~W.}\ \bibnamefont {Chin}},
  \bibinfo {author} {\bibfnamefont {S.~F.}\ \bibnamefont {Huelga}},\ and\
  \bibinfo {author} {\bibfnamefont {M.~B.}\ \bibnamefont {Plenio}},\ }\bibfield
   {title} {\bibinfo {title} {Exploiting structured environments for efficient
  energy transfer: The phonon antenna mechanism},\ }\href
  {https://doi.org/10.1021/jz400058a} {\bibfield  {journal} {\bibinfo
  {journal} {J.\ Phys.\ Chem.\ Lett.}\ }\textbf {\bibinfo {volume} {4}},\
  \bibinfo {pages} {903} (\bibinfo {year} {2013})}\BibitemShut {NoStop}%
\bibitem [{\citenamefont {Mohseni}\ \emph {et~al.}(2014)\citenamefont
  {Mohseni}, \citenamefont {Omar}, \citenamefont {Engel},\ and\ \citenamefont
  {Plenio}}]{QuEB}%
  \BibitemOpen
  \bibinfo {editor} {\bibfnamefont {M.}~\bibnamefont {Mohseni}}, \bibinfo
  {editor} {\bibfnamefont {Y.}~\bibnamefont {Omar}}, \bibinfo {editor}
  {\bibfnamefont {G.~S.}\ \bibnamefont {Engel}},\ and\ \bibinfo {editor}
  {\bibfnamefont {M.~B.}\ \bibnamefont {Plenio}},\ eds.,\ \href
  {https://doi.org/10.1017/CBO9780511863189} {\emph {\bibinfo {title} {Quantum
  Effects in Biology}}}\ (\bibinfo  {publisher} {Cambridge University Press},\
  \bibinfo {year} {2014})\BibitemShut {NoStop}%
\bibitem [{\citenamefont {Galperin}\ \emph {et~al.}(2007)\citenamefont
  {Galperin}, \citenamefont {Ratner},\ and\ \citenamefont
  {Nitzan}}]{Galperin2007}%
  \BibitemOpen
  \bibfield  {author} {\bibinfo {author} {\bibfnamefont {M.}~\bibnamefont
  {Galperin}}, \bibinfo {author} {\bibfnamefont {M.~A.}\ \bibnamefont
  {Ratner}},\ and\ \bibinfo {author} {\bibfnamefont {A.}~\bibnamefont
  {Nitzan}},\ }\bibfield  {title} {\bibinfo {title} {Molecular transport
  junctions: Vibrational effects},\ }\href
  {https://doi.org/10.1088/0953-8984/19/10/103201} {\bibfield  {journal}
  {\bibinfo  {journal} {J.\ Phys. Condens.\ Matt.}\ }\textbf {\bibinfo {volume}
  {19}},\ \bibinfo {pages} {103201} (\bibinfo {year} {2007})}\BibitemShut
  {NoStop}%
\bibitem [{\citenamefont {Chuang}\ \emph {et~al.}(2016)\citenamefont {Chuang},
  \citenamefont {Lee}, \citenamefont {Moix}, \citenamefont {Knoester},\ and\
  \citenamefont {Cao}}]{Chuang2016}%
  \BibitemOpen
  \bibfield  {author} {\bibinfo {author} {\bibfnamefont {C.}~\bibnamefont
  {Chuang}}, \bibinfo {author} {\bibfnamefont {C.~K.}\ \bibnamefont {Lee}},
  \bibinfo {author} {\bibfnamefont {J.~M.}\ \bibnamefont {Moix}}, \bibinfo
  {author} {\bibfnamefont {J.}~\bibnamefont {Knoester}},\ and\ \bibinfo
  {author} {\bibfnamefont {J.}~\bibnamefont {Cao}},\ }\bibfield  {title}
  {\bibinfo {title} {Quantum diffusion on molecular tubes: Universal scaling of
  the 1d to 2d transition},\ }\href
  {https://doi.org/10.1103/PhysRevLett.116.196803} {\bibfield  {journal}
  {\bibinfo  {journal} {Phys.\ Rev.\ Lett.}\ }\textbf {\bibinfo {volume}
  {116}},\ \bibinfo {pages} {196803} (\bibinfo {year} {2016})}\BibitemShut
  {NoStop}%
\bibitem [{\citenamefont {Pazy}\ and\ \citenamefont {Vardi}(2005)}]{Pazy2005}%
  \BibitemOpen
  \bibfield  {author} {\bibinfo {author} {\bibfnamefont {E.}~\bibnamefont
  {Pazy}}\ and\ \bibinfo {author} {\bibfnamefont {A.}~\bibnamefont {Vardi}},\
  }\bibfield  {title} {\bibinfo {title} {Holstein model and peierls instability
  in one-dimensional boson--fermion lattice gases},\ }\href
  {https://doi.org/10.1103/PhysRevA.72.033609} {\bibfield  {journal} {\bibinfo
  {journal} {Phys.\ Rev.\ A}\ }\textbf {\bibinfo {volume} {72}},\ \bibinfo
  {pages} {033609} (\bibinfo {year} {2005})}\BibitemShut {NoStop}%
\bibitem [{\citenamefont {Sch{\"a}fer}\ \emph {et~al.}(2020)\citenamefont
  {Sch{\"a}fer}, \citenamefont {Fukuhara}, \citenamefont {Sugawa},
  \citenamefont {Takasu},\ and\ \citenamefont {Takahashi}}]{Schaefer2020}%
  \BibitemOpen
  \bibfield  {author} {\bibinfo {author} {\bibfnamefont {F.}~\bibnamefont
  {Sch{\"a}fer}}, \bibinfo {author} {\bibfnamefont {T.}~\bibnamefont
  {Fukuhara}}, \bibinfo {author} {\bibfnamefont {S.}~\bibnamefont {Sugawa}},
  \bibinfo {author} {\bibfnamefont {Y.}~\bibnamefont {Takasu}},\ and\ \bibinfo
  {author} {\bibfnamefont {Y.}~\bibnamefont {Takahashi}},\ }\bibfield  {title}
  {\bibinfo {title} {Tools for quantum simulation with ultracold atoms in
  optical lattices},\ }\href {https://doi.org/10.1038/s42254-020-0195-3}
  {\bibfield  {journal} {\bibinfo  {journal} {Nat.\ Revi.\ Phys.}\ }\textbf
  {\bibinfo {volume} {2}},\ \bibinfo {pages} {411} (\bibinfo {year}
  {2020})}\BibitemShut {NoStop}%
\bibitem [{\citenamefont {Herrera}\ and\ \citenamefont
  {Krems}(2011)}]{Herrera2011}%
  \BibitemOpen
  \bibfield  {author} {\bibinfo {author} {\bibfnamefont {F.}~\bibnamefont
  {Herrera}}\ and\ \bibinfo {author} {\bibfnamefont {R.~V.}\ \bibnamefont
  {Krems}},\ }\bibfield  {title} {\bibinfo {title} {Tunable holstein model with
  cold polar molecules},\ }\href {https://doi.org/10.1103/PhysRevA.84.051401}
  {\bibfield  {journal} {\bibinfo  {journal} {Phys.\ Rev.\ A}\ }\textbf
  {\bibinfo {volume} {84}},\ \bibinfo {pages} {051401} (\bibinfo {year}
  {2011})}\BibitemShut {NoStop}%
\bibitem [{\citenamefont {Mostame}\ \emph {et~al.}(2012)\citenamefont
  {Mostame}, \citenamefont {Rebentrost}, \citenamefont {Eisfeld}, \citenamefont
  {Kerman}, \citenamefont {Tsomokos},\ and\ \citenamefont
  {Aspuru-Guzik}}]{Mostame2012}%
  \BibitemOpen
  \bibfield  {author} {\bibinfo {author} {\bibfnamefont {S.}~\bibnamefont
  {Mostame}}, \bibinfo {author} {\bibfnamefont {P.}~\bibnamefont {Rebentrost}},
  \bibinfo {author} {\bibfnamefont {A.}~\bibnamefont {Eisfeld}}, \bibinfo
  {author} {\bibfnamefont {A.~J.}\ \bibnamefont {Kerman}}, \bibinfo {author}
  {\bibfnamefont {D.~I.}\ \bibnamefont {Tsomokos}},\ and\ \bibinfo {author}
  {\bibfnamefont {A.}~\bibnamefont {Aspuru-Guzik}},\ }\bibfield  {title}
  {\bibinfo {title} {Quantum simulator of an open quantum system using
  superconducting qubits: Exciton transport in photosynthetic complexes},\
  }\href {https://doi.org/10.1088/1367-2630/14/10/105013} {\bibfield  {journal}
  {\bibinfo  {journal} {New J.\ Phys.}\ }\textbf {\bibinfo {volume} {14}},\
  \bibinfo {pages} {105013} (\bibinfo {year} {2012})}\BibitemShut {NoStop}%
\bibitem [{\citenamefont {Poto{\v{c}}nik}\ \emph {et~al.}(2018)\citenamefont
  {Poto{\v{c}}nik}, \citenamefont {Bargerbos}, \citenamefont {Schr{\"o}der},
  \citenamefont {Khan}, \citenamefont {Collodo}, \citenamefont {Gasparinetti},
  \citenamefont {Salath{\'e}}, \citenamefont {Creatore}, \citenamefont
  {Eichler}, \citenamefont {T{\"u}reci}, \citenamefont {Chin},\ and\
  \citenamefont {Wallraff}}]{Potocnik2018}%
  \BibitemOpen
  \bibfield  {author} {\bibinfo {author} {\bibfnamefont {A.}~\bibnamefont
  {Poto{\v{c}}nik}}, \bibinfo {author} {\bibfnamefont {A.}~\bibnamefont
  {Bargerbos}}, \bibinfo {author} {\bibfnamefont {F.~A. Y.~N.}\ \bibnamefont
  {Schr{\"o}der}}, \bibinfo {author} {\bibfnamefont {S.~A.}\ \bibnamefont
  {Khan}}, \bibinfo {author} {\bibfnamefont {M.~C.}\ \bibnamefont {Collodo}},
  \bibinfo {author} {\bibfnamefont {S.}~\bibnamefont {Gasparinetti}}, \bibinfo
  {author} {\bibfnamefont {Y.}~\bibnamefont {Salath{\'e}}}, \bibinfo {author}
  {\bibfnamefont {C.}~\bibnamefont {Creatore}}, \bibinfo {author}
  {\bibfnamefont {C.}~\bibnamefont {Eichler}}, \bibinfo {author} {\bibfnamefont
  {H.~E.}\ \bibnamefont {T{\"u}reci}}, \bibinfo {author} {\bibfnamefont
  {A.~W.}\ \bibnamefont {Chin}},\ and\ \bibinfo {author} {\bibfnamefont
  {A.}~\bibnamefont {Wallraff}},\ }\bibfield  {title} {\bibinfo {title}
  {Studying light-harvesting models with superconducting circuits},\ }\href
  {https://doi.org/10.1038/s41467-018-03312-x} {\bibfield  {journal} {\bibinfo
  {journal} {Nat.\ Commun.}\ }\textbf {\bibinfo {volume} {9}},\ \bibinfo
  {pages} {904} (\bibinfo {year} {2018})}\BibitemShut {NoStop}%
\bibitem [{\citenamefont {Wang}\ \emph {et~al.}(2020)\citenamefont {Wang},
  \citenamefont {Curtis}, \citenamefont {Lester}, \citenamefont {Zhang},
  \citenamefont {Gao}, \citenamefont {Freeze}, \citenamefont {Batista},
  \citenamefont {Vaccaro}, \citenamefont {Chuang}, \citenamefont {Frunzio},
  \citenamefont {Jiang}, \citenamefont {Girvin},\ and\ \citenamefont
  {Schoelkopf}}]{Wang2020}%
  \BibitemOpen
  \bibfield  {author} {\bibinfo {author} {\bibfnamefont {C.~S.}\ \bibnamefont
  {Wang}}, \bibinfo {author} {\bibfnamefont {J.~C.}\ \bibnamefont {Curtis}},
  \bibinfo {author} {\bibfnamefont {B.~J.}\ \bibnamefont {Lester}}, \bibinfo
  {author} {\bibfnamefont {Y.}~\bibnamefont {Zhang}}, \bibinfo {author}
  {\bibfnamefont {Y.~Y.}\ \bibnamefont {Gao}}, \bibinfo {author} {\bibfnamefont
  {J.}~\bibnamefont {Freeze}}, \bibinfo {author} {\bibfnamefont {V.~S.}\
  \bibnamefont {Batista}}, \bibinfo {author} {\bibfnamefont {P.~H.}\
  \bibnamefont {Vaccaro}}, \bibinfo {author} {\bibfnamefont {I.~L.}\
  \bibnamefont {Chuang}}, \bibinfo {author} {\bibfnamefont {L.}~\bibnamefont
  {Frunzio}}, \bibinfo {author} {\bibfnamefont {L.}~\bibnamefont {Jiang}},
  \bibinfo {author} {\bibfnamefont {S.~M.}\ \bibnamefont {Girvin}},\ and\
  \bibinfo {author} {\bibfnamefont {R.~J.}\ \bibnamefont {Schoelkopf}},\
  }\bibfield  {title} {\bibinfo {title} {Efficient multiphoton sampling of
  molecular vibronic spectra on a superconducting bosonic processor},\ }\href
  {https://doi.org/10.1103/PhysRevX.10.021060} {\bibfield  {journal} {\bibinfo
  {journal} {Phys.\ Rev.\ X}\ }\textbf {\bibinfo {volume} {10}},\ \bibinfo
  {pages} {021060} (\bibinfo {year} {2020})}\BibitemShut {NoStop}%
\end{thebibliography}
